\documentclass[aps,pra,floatfix,twocolumn]{revtex4} 

\usepackage{times}
\usepackage{graphicx}
\usepackage{graphics}
\usepackage{amssymb}
\usepackage{amsmath}
\usepackage{epsfig}
\usepackage{latexsym}
\usepackage{color}
\usepackage{fancyvrb}
\usepackage{subfigure}
\usepackage{multirow}
\usepackage{psfrag}
\def\figurelib{.}
\usepackage{array}

\usepackage{booktabs}

\newcolumntype{I}{!{\vrule width 1pt}}

\DefineVerbatimEnvironment{code}{Verbatim}{fontsize=\small}
\DefineVerbatimEnvironment{example}{Verbatim}{fontsize=\small}

\newcommand{\bra}[1]{\left\langle{#1}\right\vert}
\newcommand{\ket}[1]{\left\vert{#1}\right\rangle}

\newcommand{\beq}{\begin{equation}}
\newcommand{\eeq}{\end{equation}}
\newcommand{\bqa}{\begin{eqnarray}}
\newcommand{\eqa}{\end{eqnarray}}
\newcommand{\nn}{\nonumber}

\newcommand{\erf}[1]{Eq.~(\ref{#1})}

\newcommand{\eg}{\emph{e.g.},~}
\newcommand{\ie}{\emph{i.e.},~}
\newcommand{\etal}{\emph{et al.}~}

\def\idt{\mathbf{I}}

\newcommand{\specialcell}[2][c]{%
  \begin{tabular}[#1]{@{}c@{}}#2\end{tabular}}

\usepackage{theorem}

\theoremstyle{break}

\newcommand{\SJTU}{Joint Institute of UMich-SJTU and Key Laboratory of
  System Control and Information Processing (MOE),
  Shanghai Jiao Tong University, Shanghai, 200240, China}
\newcommand{\SNL}{Digital \& Quantum Information Systems,
Sandia National Laboratories, Livermore, CA 94550, USA}

\begin{document}

\title{Schedule path optimization for adiabatic quantum computing and optimization}
\author{Lishan Zeng$^1$,
Jun Zhang$^{1}$\footnote{Electronic address: \texttt{zhangjun12@sjtu.edu.cn}},
Mohan Sarovar$^{2}$\footnote{Electronic address: \texttt{mnsarov@sandia.gov}}}
\address{$^1$ \SJTU \\$^2$ \SNL}

\begin{abstract}
Adiabatic quantum computing and optimization have garnered much attention recently as possible models for achieving a quantum advantage over classical approaches to optimization and other special purpose computations. Both techniques are probabilistic in nature and the minimum gap between the ground state and first excited state of the system during evolution is a major factor in determining the success probability. In this work we investigate a strategy for increasing the minimum gap and success probability by introducing intermediate Hamiltonians that modify the evolution path between initial and final Hamiltonians. We focus on an optimization problem relevant to recent hardware implementations and present numerical evidence for the existence of a purely local intermediate Hamiltonian that achieve the optimum performance in terms of pushing the minimum gap to one of the end points of the evolution. As a part of this study we develop a convex optimization formulation of the search for optimal adiabatic schedules that makes this computation more tractable, and which may be of independent interest.
We further study the effectiveness of random intermediate Hamiltonians on the minimum gap and success probability, and empirically find that random Hamiltonians have a significant probability of increasing the success probability, but only by a modest amount.
\end{abstract}

\maketitle

\section{Introduction}
\label{sec:intro}
Controlled adiabatic evolution of a many-body quantum system is the basic ingredient for performing quantum computing via adiabatic quantum computing (AQC) \cite{Farhi:2000tw} or optimization via adiabatic quantum optimization (AQO, also referred to as quantum annealing if finite temperature effects cannot be ignored) \cite{Kadowaki:1998bm, Das:2008ko, Johnson:2011gd, Boixo:2014cg}.
The conventional approach to AQC and AQO implements a linear interpolation between an initial ($H_0$) and final ($H_1$) Hamiltonian acting on $n$ spins (or qubits):
\begin{equation}
  \label{eq:5}
H(t) = \left(1-\frac{t}{T}\right)H_0 + \frac{t}{T} H_1,
\end{equation}
where $T$ is the total time of evolution. For convenience, we hereby
reparametrize the temporal evolution by introducing a normalized time
parameter $s = t/T$ ($0\leq s \leq 1$). The system begins in the ground state of $H_0$, and the ground state of $H_1$ encodes the solution to a problem of interest. Then evolving the system according to $H(t)$ until $t=T$ and measuring permits sampling from the solution state if the system remains in the ground state throughout the evolution. The probability of this happening, or the \emph{success probability} of the adiabatic computation, relies on satisfying the following adiabatic condition ($\hbar=1$ throughout this work) \cite{Farhi:2000tw}:
\begin{equation}
\label{eq:6}
\frac{\max_{0\leq s \leq 1} |\bra{\phi_1(s)}
\frac{\partial H(s)}{\partial s}\ket{\phi_0(s)}|}
{\min_{0 \leq s \leq 1 }  \Delta^2(s)} \ll T,
\end{equation}
where $\ket{\phi_{0}(s)}$, $\ket{\phi_{1}(s)}$ are respectively the
instantaneous ground state and first excited state of $H(s)$, and
$\Delta(s)$ is the difference in energy between the first excited
state and ground state at time $s$, \ie $H(s)\ket{\phi_{0|1}(s)} =
\lambda_{0|1}(s)\ket{\phi_{0|1}(s)}$ with
$\lambda_1(s)>\lambda_0(s)$, and $\Delta(s)=\lambda_1(s) -
\lambda_0(s)$. The quantity $\Delta(s)$ is referred to as the
instantaneous \emph{gap} and its minimum value over the entire evolution as
the \emph{minimum gap} for the problem. Satisfying condition \eqref{eq:6} guarantees, with high probability, that the state of the system at the terminal time $s=1$ (\ie $t=T$)
will be the ground state of the problem Hamiltonian $H_1$ \footnote{We
  note that there have been some variations and refinements of this
  adiabatic condition in the
  literature~\cite{Marzlin:2004ki,Pati:2004ws,Schaller:2006kd,
    Jansen:2007ks,Wei:2007kr,Lidar:2009gh,Cheung:2011kb}. However, for our purposes
  this condition suffices to capture the generic behavior of success
  probability with minimum gap.}. It should be noted that the
condition in \erf{eq:6} comes from a worst-case analysis. An
alternative, \emph{local} adiabatic condition that arises directly
from the adiabatic theorem is \cite{Schaller:2006kd,Wei:2007kr}:
\begin{equation}
\label{eq:ad_cond}
\sum_{n\neq 0}\frac{|\bra{\phi_n(t)} \frac{\partial H(t)}{\partial t}
\ket{\phi_0(t)}|}{\Delta_{n0}^2(t)} \ll 1,~
\text{for all } t,
\end{equation}
where $\Delta_{n0}(t) \equiv \lambda_n(t) -\lambda_0(t)$. The sum in this condition is typically dominated by the $n=1$ term.

While the adiabatic model of quantum computation can be shown to be universal \cite{Aharonov:2007uk,Mizel:2007ci,Biamonte:2008ix}, the formulation of a fault-tolerant version of AQC remains a challenging and open problem \cite{Lid-2008,Quiroz:2012eaa,Young:2013ux,Sarovar:2013js}. Despite this deficiency, there have been substantial efforts to implement special purpose processors for optimization \cite{Johnson:2011gd, Boixo:2014cg}, and several heuristics for increasing the reliability of AQC and AQO have been developed, including encoding or increasing the degeneracy of the problem Hamiltonian \cite{Young:2013gh,Pudenz:2014gsa,Zhuang:2014jb,Pudenz:2015cl}. 

A heuristic performance enhancement strategy that has been extensively studied is the modification of the interpolation between $H_0$ and $H_1$ from linear as in \erf{eq:5} to a more general time-dependent form:
\begin{equation}
H(s) = f_0(s)H_0 + f_1(s)H_1, \nn
\end{equation}
where $f_0$ and $f_1$ are arbitrary but smoothly varying functions
satisfying boundary conditions $f_0(0)=1$, $f_0(1) = 0$ and $f_1(0)=0$,
$f_1(1)=1$. 
Several authors have shown that by tuning the funct0ions $f_{0|1}(s)$ one can increase the success probability of the adiabatic algorithm for a fixed $T$
\cite{Roland2002,Schaller:2006kd,Rezakhani:2009ix,Rezakhani:2010jr,Brif:2013vp,Hofmann:2014ct,Crosson:2014tg}. The improvement afforded by such a modification of the interpolation scheme can be understood from the intuition that it can be tuned to move slower near smaller energy gaps and faster near larger ones, provided that these regions are known \emph{a priori}. Another class of strategies, commonly known as shortcuts to adiabaticity \cite{Emmanouilidou:2000kn,Torrontegui:2013ij}, attempts to optimize $f_i(s)$ and $T$ to obtain the solution state (ground state of $H_1$) in the shortest time possible, but forfeits adiabaticity during the evolution.

In this work we study another heuristic strategy for increasing the reliability of AQC and AQO, which involves a further generalization of the adiabatic interpolation by adding additional \emph{intermediate} Hamiltonians between the two endpoints. This leads to a Hamiltonian of the form:
\begin{equation}
H(s; \{f_j(s)\}) = f_0(s) H_0 +
\sum_{j=2}^{M+1} f_j(s) H_j + f_1(s) H_1.
\label{eq:pert_interp}
\end{equation}
The $f_j$ are smoothly varying functions of $s$ with $f_0(s)$ and
$f_1(s)$ satisfying the same boundary conditions as above, and
the other $M$ schedule functions satisfying
$f_j(0)=f_j(1)=0$ for all $j \geq 2$.  The addition of
these extra Hamiltonians preserves the problem structure but
introduces the possibility of modifying the eigenvalue
distributions (and thus gaps) during time evolution. In effect, the
intermediate Hamiltonians allow the adiabatic evolution to take paths
other than a straight line from $H_0$ to $H_1$. We refer to the collection of functions $\{f_j(s)\}$ as the \emph{adiabatic schedule} and this strategy as \emph{schedule path optimization} (SPO). We will interchangeably call $H_p(s) = \sum_{j=2}^{M+1} f_j(s) H_j$ the intermediate Hamiltonian or the (adiabatically) perturbing Hamiltonian since it perturbs the straight line path. We note that early attempts to study the effects of such intermediate Hamiltonians were made in Ref. \cite{Farhi:2002vf}, and more recent investigations are in Refs. \cite{PerdomoOrtiz:2011bo,Hofmann:2014ct,Crosson:2014tg}.

We will study these intermediate Hamiltonians from the
perspective of optimization and investigate whether there exist \emph{local}
intermediate Hamiltonians that increase the success probability of computations.  We will consider only local, or single-qubit, intermediate Hamiltonians for
two reasons: (i) these are likely the most experimentally feasible; and (ii) the number of intermediate terms is restricted to $M=3n$ at most, for an $n$-qubit problem, and thus the search for optimal perturbing Hamiltonians is tractable.  We show that the problem of finding good intermediate Hamiltonians can be approximated by a convex optimization, which considerably increases the solution efficiency. However, even with a convex optimization approach, explicit solution of the optimal schedule path quickly becomes infeasible as $n$ increases. Therefore, we also study the likelihood of random
local intermediate Hamiltonians enhancing the success of computations. In
addition to AQC and AQO, the approach considered can inform other quantum adiabatic protocols such as adiabatic state and
population transfer \cite{Greentree:2004di}. In such cases, one may be
interested in systems with moderate state space dimension, and the
approach we outline can be used to constructively find intermediate Hamiltonians that increase the probability of success.

The remainder of the paper is organized as follows. Section \ref{sec:prelim} introduces the class of adiabatic evolution problems we will focus on, and explicitly formulates the optimization problem of interest. Then in section \ref{sec:optimal} we apply a convex approximation to the optimization of the intermediate Hamiltonian, present solutions to the problem, and study their properties. In section \ref{sec:random} we consider the strategy of introducing random intermediate Hamiltonians and study its effectiveness. Finally, section \ref{sec:disc} summarizes the findings and discusses implications.

\section{Preliminaries}
\label{sec:prelim}
For our investigations we focus on adiabatic
interpolations that solve quadratic unconstrained binary optimization (QUBO)
problems. This NP-hard problem \cite{Pardalos:1992hi} is phrased as a minimization of a sum clause of binary variables, for example:
\begin{equation*}
\begin{aligned}
  &\min_{x_1,x_2} \ ax_1 + bx_2 + c x_1 x_2  \\
  \textrm{s. t.} \  & x_1, x_2 \in \{0,1\}.
\end{aligned}
\end{equation*}
A quantum approach to solve such a problem with $n$ variables is to map the solution to the ground state of the final Hamiltonian of an $n$-qubit system, and then let the quantum system adiabatically evolve from the ground state of a readily prepared initial Hamiltonian to the ground state of that final Hamiltonian. In particular, we can choose the final Hamiltonian as
\begin{equation}
\label{eq:qubo_prob}
H_1 = \sum_i h_i \sigma_z^i + \sum_{i,j} J_{ij} \sigma_z^i \sigma_z^j,
\end{equation}
where the eigenstates of $\sigma_z^i$ correspond to the binary values
of variable $x_i$, and the Hamiltonian coefficients $h_i$ and $J_{ij}$
are functions of the problem coefficients ($a,b,c$ in the example
above). Note that we suppress tensor product symbols and identities
  when denoting $n$-qubit Pauli matrices, and superscripts indicate the
  qubits on which there is a non-trivial action; \ie $\sigma^2_x \equiv
  I \otimes \sigma_x \otimes I \otimes \cdots \otimes I$. The initial
Hamiltonian can be set in the standard form
\begin{equation}
  \label{eq:8}
H_0 =\sum_i \sigma_x^i.
\end{equation}
Although it
is not possible to transform any classical computation to a QUBO problem these problems are particularly relevant since recently developed experimental devices attempt to solve problems from this class \cite{Johnson:2011gd, Boixo:2014cg}.

\subsection{Problem specification}
In order to assess the effectiveness of local intermediate Hamiltonians,
we must define a performance metric. One would ideally use the probability of success
$p_{\rm succ}(T) \equiv |\langle{\phi_0(T)}|\psi(T)\rangle|^2$, where
$\ket{\phi_0(T)}$ is the ground state of $H_1$ and
$\ket{\psi(T)}$ is the actual achieved state at the final time.
However, for the ease of optimization formulation,
we instead use a proxy for the success probability, namely,
the minimum gap $\Delta_\textrm{min} = \min_s
\Delta(s)$. This
enables us to approximate the intermediate Hamiltonian optimization problem by a convex optimization, which
facilitates numerical investigation of intermediate size qubit systems. In Sec.
\ref{sec:optimal} we will confirm that the optimal intermediate
Hamiltonians identified by this proxy cost function do indeed enhance the probability of success for large T in most cases.

Now we are in a position to formulate the optimization problem.
Given an adiabatic problem specification with $H_0$, $H_1$, and a
set of $M$ local, intermediate Hamiltonians $\{H_j\}_{j=2}^{M+1}$, the
optimal schedule path is defined as the one that solves:
\begin{equation}
  \label{eq:1}
  \begin{aligned}
  &\max_{\{f_j(s)\}_{j=2}^{M+1}} \quad
\min_{0\leq s \leq 1} ~~\Delta(s; \{f_j(s)\})  \\
\textrm{s. t.}\quad &  f_j(0)=f_j(1)=0,\quad \forall\; j\geq 2, \\
& |f_j(s)|\le f_B, \quad \forall\; j\geq 2,\\
& |\dot{f}_j(s)| < \varepsilon, \quad \forall\; j \geq 2,
  \end{aligned}
\end{equation}
where $\Delta(s; \{f_j(s)\})$ is the gap of $H(s; \{f_j(s)\})$. In other words, the optimal schedule path maximizes, over the set of interpolation functions, the minimum gap on the entire time duration. The first set of constraints forces the intermediate Hamiltonians to be zero at the initial and terminal time, the second
set bounds the amplitude of the intermediate terms, and the last set
ensures that the control functions are smooth ($\varepsilon$ is a
small positive constant).  We fix $f_0$ and $f_1$ as $f_0(s)=1-s$ and
$f_1(s)=s$ for simplicity and also to isolate the effects of the intermediate Hamiltonians
from effects of varying the interpolation velocity.

The minimax optimization above is difficult to tackle in its
continuous form. Therefore we first discretize time to obtain a more
tractable formulation. Divide the total time duration $s\in[0,1]$ into
$N$ equal intervals each of length $\Delta s=\frac{1}{N}$. Within the
interval $[i\Delta s,(i+1)\Delta s)$, assume that the schedule
functions $f_j(s)$ take a constant value $f_j(i) \equiv f_j(i\Delta
s)$, where $i=0$, $\cdots$, $N-1$. In particular, $f_j(0)=f_j(N)=0$. When $N$ is large enough, this piecewise constant function approximates the continuous function well. We then define a
$M(N-1)$ dimensional discrete optimization variable
\begin{equation}
A = [f_2(1), f_2(2), ...., f_{M+1}(N-2), f_{M+1}(N-1)] \nn
\end{equation}
and the discrete optimization problem can  be written as
\begin{equation}
  \label{eq:3}
  \begin{aligned}
 &\max_{A}\quad \min_{1\leq i \leq N-1}\quad\Delta_i(A) \\
\text{s. t. }&|f_j(i)|\leq f_{B},~i=1, \cdots, N-1,\\
& |f_j(i+1)-f_j(i)|\leq\varepsilon \Delta s,~i=0, \cdots, N-1,
  \end{aligned}
\end{equation}
where $j=2$, $\cdots$, $M+1$. The cost function $\Delta_i(A)$ is
the energy gap between the ground and the first excited states at
$s=i\Delta s$ when the particular adiabatic schedule defined by $A$ is
used.

For small number of qubits, this discrete optimization can be solved
efficiently using standard sequential quadratic programming methods.  In particular, we applied the Matlab Optimization
Toolbox to solve for the (not necessarily globally) optimal intermediate Hamiltonian for a 10-qubit QUBO problem in 82.9 hours on an ordinary
desktop computer. However, when the number of qubits increases further,
the numerical solution quickly exceeds the computational power of available
computers. To remedy it, we will formulate a convex
optimization approximation to Eq.~\eqref{eq:3} in the next section.

Before proceeding we note that this method of gap optimization cannot
be utilized to increase success probability if the minimum gap for the
problem occurs at the endpoints of the interpolation (\ie at $s=0$ or $1$).
However, most QUBO problems do not have minimum gaps at $s=0$ or $1$ under linear interpolation.

\section{Optimal intermediate Hamiltonians from convex optimization}
\label{sec:optimal}
We now approximate the search for optimal adiabatic schedules by a
convex optimization formulation motivated by the methods in Ref. \cite{Men:2010bk}. Convexity ensures that any local optimum is a global optimum and a convex problem can be solved efficiently with mature numerical methods \cite{boyd:04}. Thus this approximation technique makes it possible to numerically search for optimal adiabatic schedules for larger AQC problems; \eg QUBO problems up to $22$ qubits readily.
 
\subsection{Approximate convex optimization formulation}
Denote the Hamiltonian at time instant $i\Delta s$ as $H(i)$, and its
eigenvalues and eigenvectors as $\lambda_k(i)$ and $u_k(i)$, \ie
\begin{equation*}
  H(i) u_k(i)=\lambda_k(i) u_k(i),~ i=0, \cdots, N.
\end{equation*}
For brevity of notation, we drop the dependency on the optimization
variable $A$. Since $H(i)$ is Hermitian, we assume that $u_k(i)$ are
chosen as an orthonormal basis and $\lambda_k(i)$ are arranged in
ascending order. Consider an upper bound $\epsilon_0(i)$ for
$\lambda_0(i)$ and a lower bound $\epsilon_1(i)$ for $\lambda_1(i)$.
The discrete optimization problem Eq.~\eqref{eq:3} can be written as
\begin{equation}
  \label{eq:2}
  \begin{aligned}
 &\max_{A, \epsilon_0(i), \epsilon_1(i)}\quad \min_{1\leq i \leq N-1}\quad
\epsilon_1(i)-\epsilon_0(i) \\
\text{s. t. }\ 
& \lambda_0(i)\leq \epsilon_0(i),~i=1, \cdots, N-1, \\
&  \lambda_1(i) \geq \epsilon_1(i),~i=1, \cdots, N-1, \\
&|f_j(i)|\leq f_{B},~i=1, \cdots, N-1,\\
& |f_j(i+1)-f_j(i)|\leq\varepsilon \Delta s,~i=0, \cdots, N-1,
  \end{aligned}
\end{equation}
where $j=2$, $\cdots$, $M+1$. The first two constraints in
\eqref{eq:2}, \ie the bounds on $\lambda_0(i)$ and $\lambda_1(i)$, can
also be written as
\begin{equation}
\label{equ:sede}
\begin{aligned}
&{\Phi_0^\dag(i)} \left(H(i)-\epsilon_0(i)\idt\right)
\Phi_0(i)\leq 0,\\
&{\Phi_1^\dag(i)} \left(H(i)-\epsilon_1(i)\idt\right)
\Phi_1(i)\succeq 0,\\
\end{aligned}
\end{equation}
where
\begin{eqnarray}
  \label{eq:4}
\Phi_0(i) &=&[u_0(i)], \\
  \label{eq:11}
\Phi_1 (i)&=&[u_1(i), \cdots, u_{2^n-1}(i)].
\end{eqnarray}
Here $\succeq$ is the L\"{o}wner partial ordering on positive
semidefinite matrices, \ie $A\succeq B$ if and only if $A-B$ is
positive semidefinite.
 
We can use Eq.~\eqref{equ:sede} to substitute the first two inequality
constraints in \eqref{eq:2} and obtain an equivalent optimization
problem. This is still a nonconvex problem, because the
constraints~\eqref{equ:sede} are nonconvex.  However, we can modify
the formulation slightly by using an iterative procedure where the
eigenvectors $u_k(i)$ from the previous iteration are used to form the
matrices $\Phi_0(i)$ and $\Phi_1(i)$ for the current iteration. This way, $\Phi_0(i)$ and $\Phi_1(i)$ become constant matrices and do
not depend on the current Hamiltonian.  We then obtain a convex
optimization problem since both the cost function and constrains are
convex.  Provided that the iteration step size is small enough, it can
well approximate the original adiabatic schedule search problem (more about this approximation later in this section).
 
When dealing with a system with $n$ qubits, the matrix $\Phi_1(i)$ has
$2^n-1$ columns, which may demand excessive computational power for
large value of $n$. We thus drop the eigenvectors in $\Phi_1(i)$ that
correspond to higher energies to reduce the problem
size~\cite{Men:2010bk}, that is, we truncate
\begin{equation}
\label{equ:redpsi}
\Phi_1(i)=[u_1(i), \cdots, u_p(i)], ~~~~ p \ll 2^n-1.
\end{equation}
In our numerical studies of systems up to $n=22$, we have observed
that $p=5$ suffices to yield good results.
 
Now after these two approximations, we have the following reduced
approximative convex optimization problem formulation:
\begin{equation}
\label{equ:convex3}
\begin{aligned}
 &\max_{A, \epsilon_0(i), \epsilon_1(i)}\quad \min_{1\leq i \leq N-1}\quad
\epsilon_1(i)-\epsilon_0(i) \\
\text{s. t. }\ 
&{\Phi_0^\dag(i)} \left(H(i)-\epsilon_0(i)\idt\right)
\Phi_0(i)\leq 0,~i=1, \cdots, N-1,\\
&{\Phi_1^\dag(i)} \left(H(i)-\epsilon_1(i)\idt\right)
\Phi_1(i)\succeq 0,~i=1, \cdots, N-1, \\
&|f_j(i)|\leq f_{B},~i=1, \cdots, N-1,\\
& |f_j(i+1)-f_j(i)|\leq\varepsilon \Delta s,~i=0, \cdots, N-1,
\end{aligned}
\end{equation}
where $\Phi_0^\dag(i)$ is defined in Eq.~\eqref{eq:4} and
$\Phi_1^\dag(i)$ in Eq.~\eqref{equ:redpsi}.  The complete algorithm to
solve for the optimal adiabatic schedule can be summarized as follows:
\begin{description}
\item[\textbf{Step} 1] Set the initial guess $A=0$ and let $\hat{A}=A$
  in the first iteration. Set the values of two small positive
  constants $\xi$ and $\eta$. Calculate the minimum gap $\Delta_0$
  with linear interpolation.
\item[\textbf{Step} 2] Compute the matrices $\Phi_0(i; \hat{A})$ and
  $\Phi_1(i; \hat{A})$. Solve the convex optimization
  problem~\eqref{equ:convex3} with the additional convex constraint
  $\|A-\hat{A}\|\leq \eta$ to ensure the resulting optimal solution
  $A^*$ close to $\hat{A}$.
\item[\textbf{Step} 3] If $\|\min \limits_{1\le i\le N-1} \Delta(i; A^*)
  -\min \limits_{i=0, N} \Delta(i; A)\|\leq\xi$, stop the algorithm
  and return the optimal solution $A^*$; otherwise, let $\hat{A}=A^*$
  and go back to \textbf{Step 2}.
\end{description}
The additional convex constraint specified in Step 2 of this algorithm aids in restricting the change in the optimization variable $A$ between iterations, which is necessary for the eigenvectors $u_k(i)$ of previous iteration to be a good approximation to the eigenvectors of the Hamiltonian for the current iteration. This is obviously not a sufficient condition to ensure the validity of this approximation all the time, but in practice we have found that the above algorithm performs well when this condition is imposed. 

\subsection{Optimal adiabatic schedules}
\label{sec:optim}
We now present the results of our search for optimal local intermediate Hamiltonians for the QUBO problem. We investigate a variety of QUBO problems, ranging from $3$ to $22$ qubits in size. For $n\leq 10$ qubits we directly solve the minimax optimization problem formulated in \erf{eq:3}; however, for larger problem sizes we solve the convex approximation  developed above. For each problem size, we do not assume any special structure for the local terms or the couplings, but rather, we generate random problem Hamiltonians by sampling each $h_i$ and $J_{ij}$ in \erf{eq:qubo_prob} from a uniform distribution on the interval $[-1,1]$. In the following, we present results from a 19-qubit and a 22-qubit system. These results are representative of the general properties of solutions to all the problems we investigated.

The initial Hamiltonian is set to be $H_0=\sum_i \sigma_x^i$, and thus the gap at the initial time is $2$, which is the cost of flipping the phase of any one qubit. For the two examples considered here, we choose
the final Hamiltonian $H_1$ to also have a gap around $2$ and the minimum gap under linear interpolation to be located away from the two endpoints. These choices are made to demonstrate the effect of optimal adiabatic schedules more effectively. The intermediate Hamiltonian takes the form
\beq
H_p^{\rm QUBO}(s) = \sum_{j=1}^n \left(f_{2j}(s) \sigma_x^j + f_{2j+1}(s) \sigma_z^j\right) ,
\label{eq:qubo_Hp}
\eeq 
and hence $M=2n$. We fix the transverse field in the $x$ direction and ignore local $\sigma_y$ perturbations in $H_0$ and $H_p$. This is because the sum of a local $\sigma_y$ and $\sigma_x$ term at any $s$ is still a net local field in the transverse direction orthogonal to the problem Hamiltonian and thus a local $\sigma_y$ term does not have any additional effects.

We also divide the time interval $[0, 1]$ into $N=50$ equal steps, and set the maximum allowed changing rate of schedule functions as $\varepsilon=2.5$. Numerical studies reveal that varying the discretization number $N$ of the time interval has little effect on optimization results as long as its reasonably large, and we shall explore the effect of varying $\varepsilon$ later. Finally, the maximum value of the intermediate terms $f_B$ is set to be the maximal value of $|h_i|$ and $|J_{ij}|$, ensuring that the energy of the local terms in the intermediate Hamiltonian is never excessive compared to the terms in the problem Hamiltonian.
\begin{figure}[t]
\centering \psfrag{Original}[][][0.9]{Linear}
\psfrag{Optimization}[][][0.9]{~~~SPO} \psfrag{y}[][][0.9]{Gap}
\psfrag{x}[][][0.9]{Time (s)} \psfrag{z}[][][0.9]{}
\psfrag{0}[][][0.8]{0} \psfrag{0.2}[][][0.8]{0.2} \psfrag{0.4}[][][0.8]{0.4}
\psfrag{0.6}[][][0.8]{0.6} \psfrag{0.8}[][][0.8]{0.8} \psfrag{1}[][][0.8]{1}
\psfrag{0.1}[][][0.8]{0.1} \psfrag{0.3}[][][0.8]{0.3}
\psfrag{-0.1}[][][0.8]{-0.1} \psfrag{-0.3}[][][0.8]{-0.3}
\psfrag{-0.2}[][][0.8]{-0.2} \psfrag{-0.4}[][][0.8]{-0.4}
\psfrag{-1}[][][0.8]{-1} \psfrag{0.5}[][][0.8]{0.5} \psfrag{-0.5}[][][0.8]{-0.5}
\psfrag{1.5}[][][0.8]{1.5} \psfrag{2.5}[][][0.8]{2.5} \psfrag{3}[][][0.8]{3}
\psfrag{2}[][][0.8]{2}
\includegraphics[width=0.69\hsize]{\figurelib/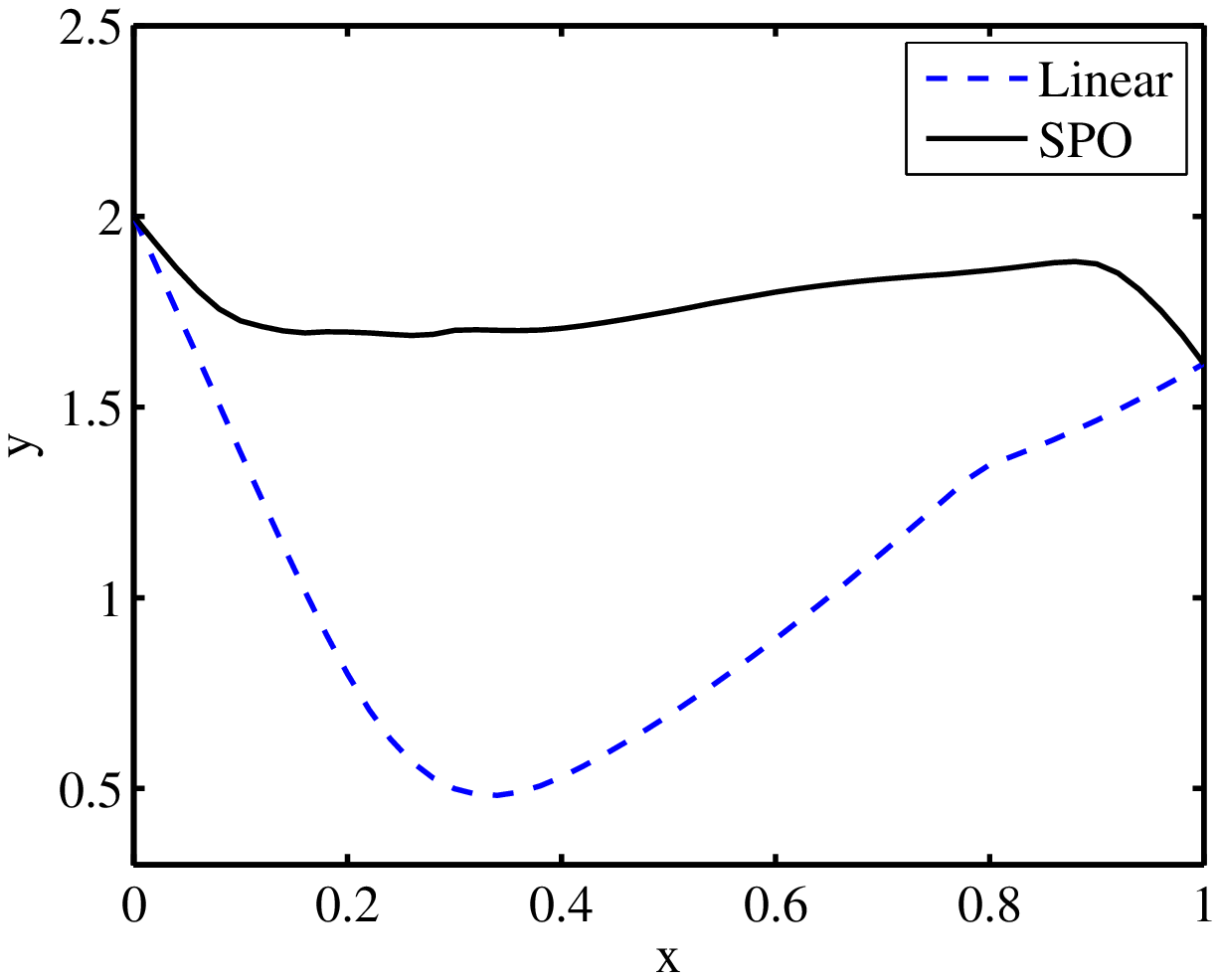}\\
\includegraphics[width=0.71\hsize]{\figurelib/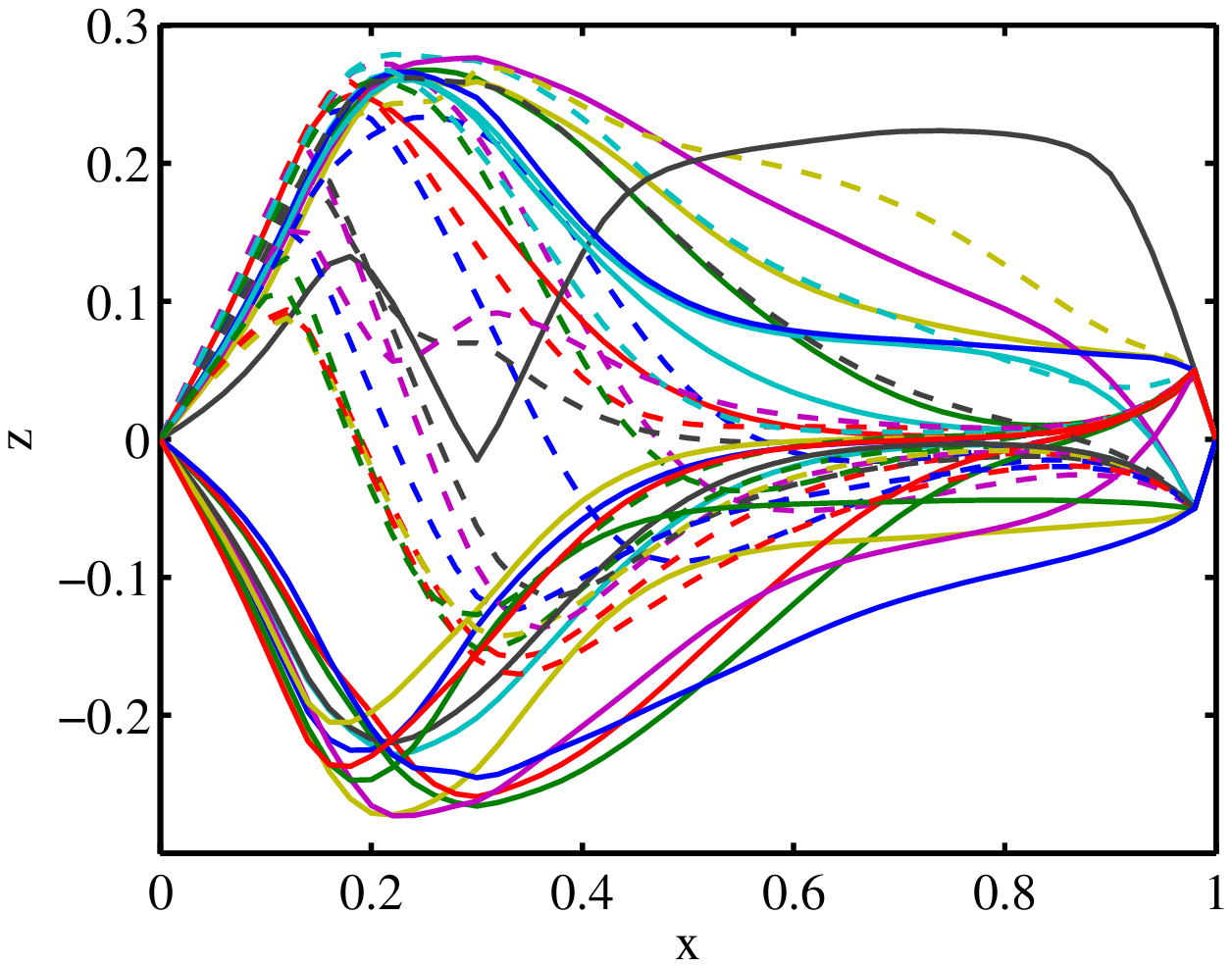}
\caption{(Color online) Optimized adiabatic schedules for a $19$-qubit AQC problem.
  Top: Energy gap, $\Delta(s)$ for linear interpolation according to
  Eq.~\eqref{eq:5} (blue dashed), and after SPO (black solid). Bottom: The $M=38$ schedule functions, $f_j(s)$, for the optimized adiabatic schedule. The dashed lines show the time-dependent coefficients of local $\sigma_x$ terms and the solid lines show the coefficients of local $\sigma_z$ terms.}
\label{fig:qubit19}
\end{figure}

\begin{figure}[t]
\centering \psfrag{Linear}[][][0.9][0.9]{Linear}
\psfrag{SPO}[][][0.9][0.9]{SPO} \psfrag{y}[][][0.9]{Gap}
\psfrag{x}[][][0.9]{Time (s)} \psfrag{z}[][][0.9]{}
\psfrag{0}[][][0.8]{0} \psfrag{0.2}[][][0.8]{0.2} \psfrag{0.4}[][][0.8]{0.4}
\psfrag{0.6}[][][0.8]{0.6} \psfrag{0.8}[][][0.8]{0.8} \psfrag{1}[][][0.8]{1}
\psfrag{-0.2}[][][0.8]{-0.2} \psfrag{-0.4}[][][0.8]{-0.4}
\psfrag{-0.6}[][][0.8]{-0.6} \psfrag{-0.8}[][][0.8]{-0.8}
\psfrag{-1}[][][0.8]{-1} \psfrag{0.5}[][][0.8]{0.5} \psfrag{-0.5}[][][0.8]{-0.5}
\psfrag{1.5}[][][0.8]{1.5} \psfrag{2.5}[][][0.8]{2.5} \psfrag{3}[][][0.8]{3}
\psfrag{2}[][][0.8]{2}
\includegraphics[width=0.7\hsize]{\figurelib/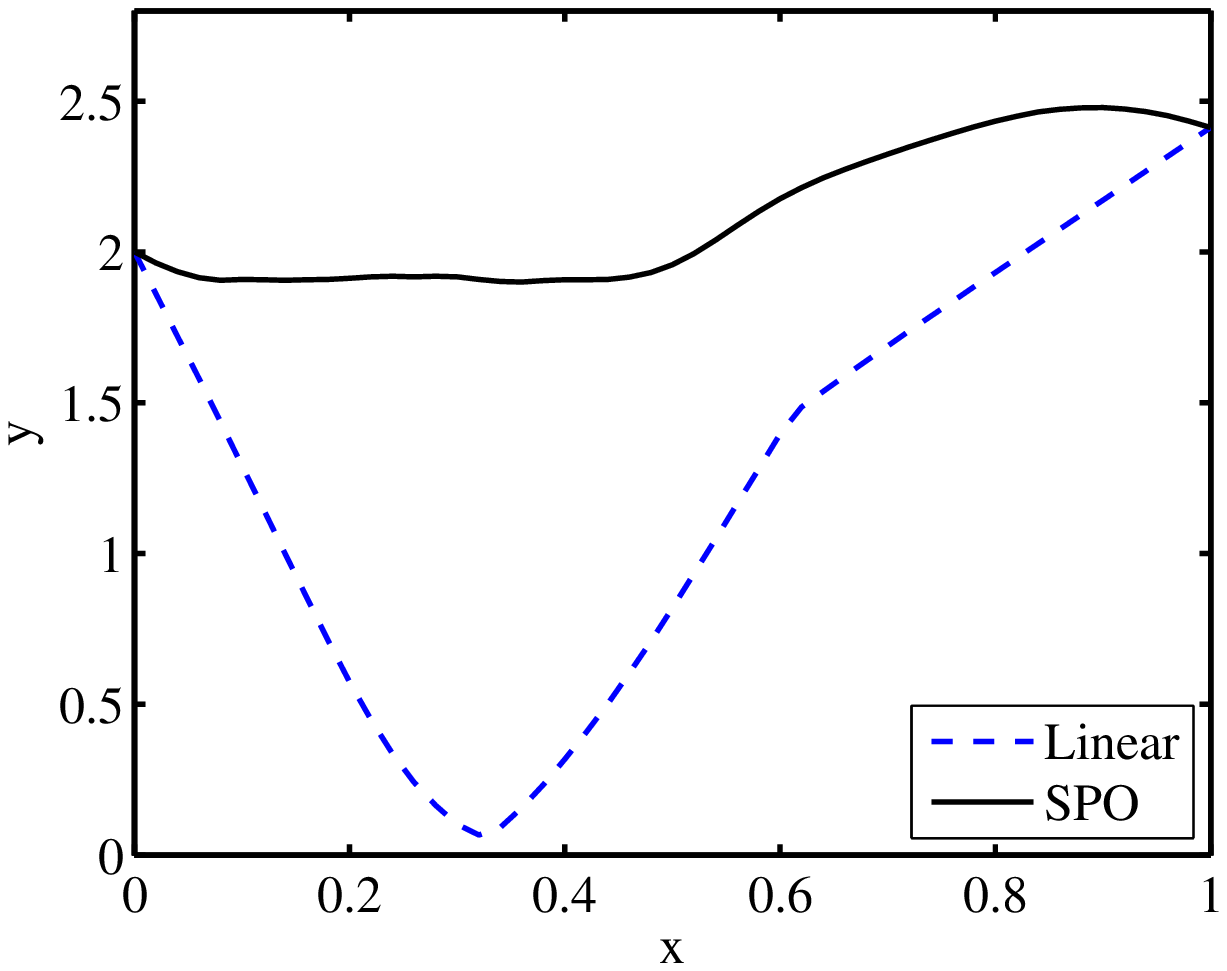}
\includegraphics[width=0.7\hsize]{\figurelib/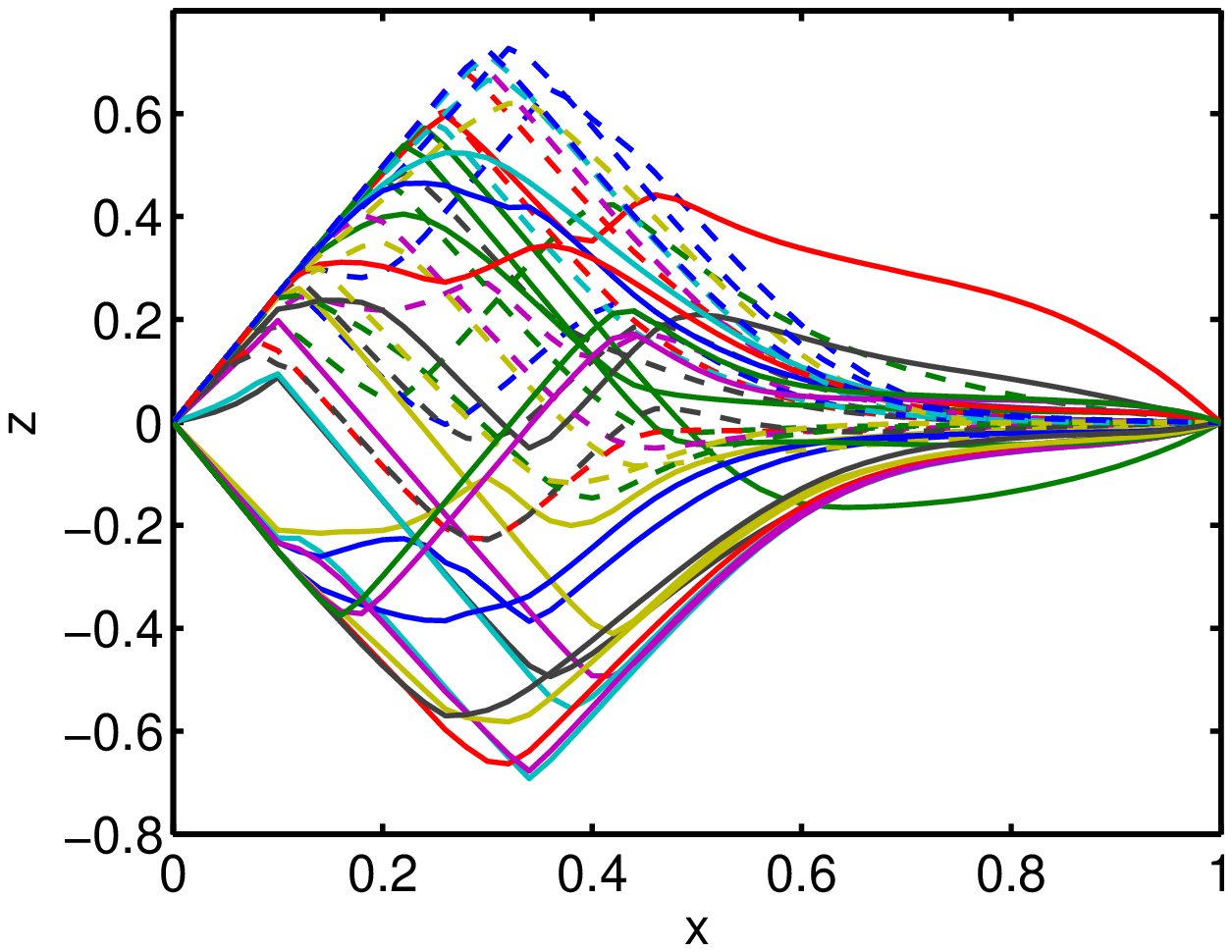}
\caption{
(Color online) Optimized adiabatic schedules for a $22$-qubit AQC problem.
  Top: Energy gap, $\Delta(s)$ for linear interpolation according to
  Eq.~\eqref{eq:5} (blue dashed), and after SPO (black solid). Bottom: The $M=38$ schedule functions, $f_j(s)$, for the optimized adiabatic schedule. The dashed lines show the time-dependent coefficients of local $\sigma_x$ terms and the solid lines show the coefficients of local $\sigma_z$ terms.}
\label{fig:qubit22}
\end{figure}

The results of solving \erf{equ:convex3} for a 19-qubit
QUBO instance are shown in Fig.~\ref{fig:qubit19}. After 2
iterations and $2.2$ hours on a single processor we obtain adiabatic schedules that can push the minimum gap to the terminal time. Fig.~\ref{fig:qubit22} shows the results for a 22-qubit instance after 5 iterations and $112$ hours. In this case, the minimum gap increases with each iteration of the convex optimization as the adiabatic schedule is refined further. Most of the computing time in both cases is devoted to calculating eigenvectors and eigenvalues as opposed to performing the convex optimization.

A striking feature present in both solutions in Figs. \ref{fig:qubit19} and \ref{fig:qubit22} is that for the (locally) optimal adiabatic schedule, the minimum gap is equal to, or almost equal to, the minimum of the initial and final gaps. We refer to this as the best-case performance (\ie $\min_s \Delta(s) = \min \{\Delta(0), \Delta(1)\}$) since the minimum gap has been pushed to one of the endpoints, where $H_p=0$. This is seen to be a generic feature of the optimal adiabatic schedules for all the QUBO problems we studied. In the non-convex formulation, which is solvable for $n\sim 10$ qubits, the optimal adiabatic schedule achieved this best-case performance for all the instances that we solved. For the convex approximation, which is solvable for larger $n$, this best-case performance is approached with increasing iterations.

We have not been able to discern any pattern in the form of the optimal schedule functions other than that some of them generally peak near the region of minimum gap for the linear interpolation and decay (mostly smoothly) to zero, whereas some others go across zero near this region.

\subsection{Effect of optimization parameters}
\begin{figure}[tb]
\centering
\psfrag{e}[][][0.8]{$\epsilon$}
\psfrag{g}[][][0.8]{Gap}
\psfrag{s}[][][0.8]{Time (s)}
\psfrag{mg}[][][0.8]{Minimum gap}
\includegraphics[width=0.9\hsize]{\figurelib/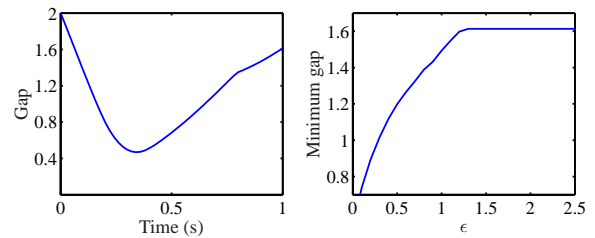}
\caption{The effect of $\varepsilon$ on the optimization performance for the 19-qubit example discussed in Fig.~\ref{fig:qubit19}. Left: the gap as a function of $s$ for linear interpolation; Right: the minimum gap achieved by the optimization as a function of $\varepsilon$.} \label{fig:ep19}
\end{figure}

The primary free parameter in all formulations of the gap optimization (Eqs. \eqref{eq:1}, \eqref{eq:3}, and \eqref{equ:convex3}) is the bound on the rate of change of the schedule functions, $\varepsilon$. Allowing for large slew rates increases $\partial H/\partial s$, which in turn increases the runtime of the adiabatic algorithm, as illustrated by \erf{eq:6}. At the same time, the larger this parameter is the more freedom there is in varying adiabatic schedule functions, and this can enable greater tuning of the system minimum gap.

This is evidenced in Fig. \ref{fig:ep19} where we show the minimum gap achieved by SPO for the 19-qubit problem presented in Fig.~\ref{fig:qubit19}, as a function of $\varepsilon$. The optimization is performed using the convex optimization formulation \eqref{equ:convex3}. We see that for small $\varepsilon$ the maximum minimum gap achieved by the optimization is modest, and then as the value of $\varepsilon$ increases the minimum gap saturates to the best-case performance (\ie $\min_s \Delta(s) = \min \{\Delta(0), \Delta(1)\}$). However, the minimum value of $\varepsilon$ to achieve this best-case performance is problem dependent and cannot be determined \emph{a priori}.

\subsection{Relation to AQC success probability}
\label{sec:succ}
\begin{table*}[htbp]
\footnotesize
  \centering
    \begin{tabular}{cIcc|cc|ccIcc|cc|ccIcc} \hline
    \toprule
    \multirow{3}{*}{Success probability} & \multicolumn{6}{cI}{10-qubit}                        & \multicolumn{6}{cI}{12-qubit}                        & \multicolumn{2}{c}{19-qubit} \\ \cline{2-15}
    \cline{2-15}
    &  \multicolumn{2}{c|}{Case 1} & \multicolumn{2}{c|}{Case 2} & \multicolumn{2}{cI}{Case 3} & \multicolumn{2}{c|}{Case 4} & \multicolumn{2}{c|}{Case 5} & \multicolumn{2}{cI}{Case 6} & \multicolumn{2}{c}{Case 7} \\ \cline{2-15}
   & linear & SPO  & linear & SPO  & linear & SPO  & linear & SPO  & linear & SPO  & linear & SPO  & linear & SPO \\ \hline
     $T=5$   &0.2441  & 0.1838  & 0.1660  & 0.5418  & 0.2846  & 0.6856  & 0.1341  & 0.3182  & 0.1398  & 0.1861  & 0.1846  & 0.8748  & 0.1011  & 0.2567  \\
    $T=10$  &0.3608  & 0.8746  & 0.3338  & 0.9223  & 0.5857  & 0.9761  & 0.2738  & 0.9553  & 0.2013  & 0.9203  & 0.3584  & 0.9604  & 0.3898  & 0.8097  \\
    $T=20$  &0.4259  & 0.9885  & 0.5810  & 0.9843  & 0.8194  & 0.9903  & 0.3907  & 0.9942  & 0.2583  & 0.9702  & 0.5822  & 0.9872  & 0.7506  & 0.9888  \\
    $T=40$  &0.4714  & 0.9947  & 0.8282  & 0.9981  & 0.9633  & 0.9959  & 0.4561  & 0.9984  & 0.5054  & 0.9964  & 0.8152  & 0.9950  &  0.9619& 0.9957 \\ \hline
   Minimum gap&0.1106&1.6704&0.2989&1.6549&0.4728&1.6659&0.1381&1.5491&0.2245&1.5862&0.4231&2.000&0.4823&1.6135\\
    Time min. gap occurs &$s=0.50$ &$s=1$&$s=0.44$ &$s=1$&$s=0.38$ &$s=1$&$s=0.44$ &$s=1$&$s=0.50$ &$s=1$&$s=0.38$ &$s=0$&$s=0.34$ &$s=1$\\
    \bottomrule \hline
    \end{tabular}%
  \caption{Success probability for different running times $T$, with linear and SPO interpolations, for seven problem Hamiltonians. For each case, the minimum gaps under linear interpolation and the optimized schedule are shown, as well as the time when the minimum gap occurs.}
  \label{tab:succ_prob}%
\end{table*}%

We have demonstrated that local intermediate Hamiltonians can increase the minimum gap of QUBO problems; however, as mentioned in the Introduction, the minimum gap is a proxy for the success probability of an AQC. Increasing the minimum gap is likely to increase success probability (given a fixed running time $T$), but the relationship between these quantities has been shown to be non-monotonic in some cases, \eg \cite{Cullimore:2012kk}. 

We now examine the relation between the success probability for the optimized and linear interpolation adiabatic schedules. Table \ref{tab:succ_prob} lists the success probabilities for several problem instances and total evolution times, with the optimized adiabatic schedule as determined by maximizing the minimum gap. We see that in most cases, increasing the minimum gap also increases the probability of success, with one exception when the running time is short (case 1, $T=5$). We determined that this exception is an example of where the increase in minimum gap is not sufficient to compensate for the increased variation produced by SPO. That is, the optimized adiabatic schedules typically have greater variation in the Hamiltonian, $|\partial H/\partial s|$, than the linear interpolation, and for short $T$, the increase in the minimum gap may not be enough to compensate for this increase in variation. This is the reason that as $T$ is increased the success probability under SPO exceeds that of linear interpolation. However, in the other extreme, where $T$ tends to infinity, there is also no advantage to SPO since for arbitrarily long running times the minimum gap is inconsequential (we do not show data for such long running times in Table \ref{tab:succ_prob} but have noticed this effect in the numerical studies). Of course this regime is impractical since other factors and timescales usually set a maximum affordable running time.

\subsection{Hard QUBO instances}
\label{sec:hardest}
\begin{figure}
\centering
 \psfrag{Gap}[][][1]{Gap}
  \psfrag{s}[][][1]{Time (s)}
\includegraphics[width=.63\hsize]{\figurelib/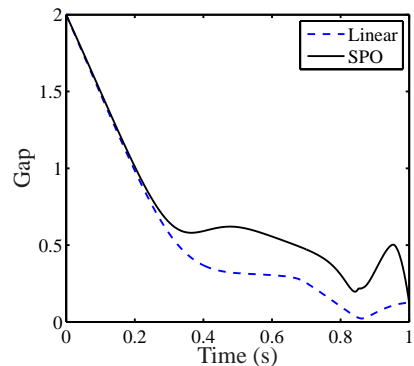}\\
{(a) Hard case 1: $\Delta^0_{\rm min}=0.0222$, $s_{\rm min}\sim 0.86$, $p^0_{\rm succ}=0.0035$.}\\
\includegraphics[width=.63\hsize]{\figurelib/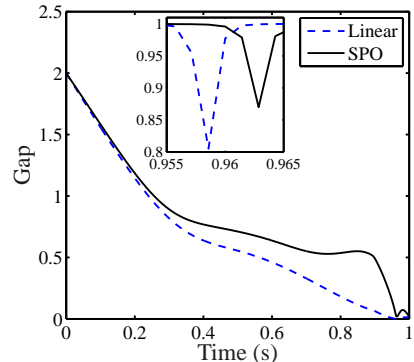}\\
{(b) Hard case 2: $\Delta^0_{\rm min}=0.0006$, $s_{\rm min}\sim 0.96$, $p^0_{\rm succ}=0.0066$.}\\
\includegraphics[width=.63\hsize]{\figurelib/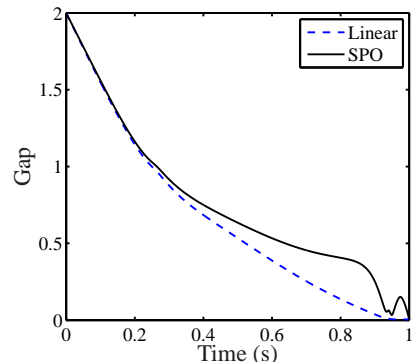}\\
{(c) Hard case 3: $\Delta^0_{\rm min}=0.0044$, $s_{\rm min}\sim 0.96$, $p^0_{\rm succ}=0.0008$}.
\caption{(Color online) The time-dependent gap for the three hardest instances of QUBO we found, with linear and SPO interpolations. In the above, $\Delta^0_{\rm min}$ is the minimum gap under linear interpolation that occurs at $s_{\rm min}$, and $p^0_{\rm succ}$ is the probability of success under linear interpolation with $T=10$. The inset in (b) shows the overlap of neighboring ground states, $|\langle u_0(i)|u_0(i+1) \rangle|$. The sudden drop in this quantity around $s_{\rm min}$ indicates a finite-size analog of a phase transition. We do not show this overlap/fidelity for the other problem instances since they do not possess a phase transition, and hence do not show a significant change in this quantity when crossing $s_{\rm min}$.}
\label{fig:hardpermy}
\end{figure}

In this subsection we study very hard instances of QUBO, and whether SPO can improve performance for these cases. To do so, we generated 180,000 random instances of $10$-qubit QUBO problems and selected the 30 instances with the smallest success probability for linear interpolation between $H_0$ and $H_1$ with $T=10$. For conciseness, in the following we only present data for the three hardest from these 30 instances (ones with the lowest success probability under linear interpolation), but the conclusions we draw are based on studying the large set of all 30 hard instances. The SPO optimizations were performed using the convex approximation in \erf{equ:convex3} with $\epsilon=2.5$. 

Fig.~\ref{fig:hardpermy} shows the time-dependent gap for the three hardest instances with linear and SPO interpolations. These three cases exhibit two features that are common to all of the 30 hard instances we investigated. The first feature is that with linear interpolation the minimum gap always occurs near $s=1$ for the hard problems. The second feature is that with the optimized schedule, the minimum gap can always be pushed out to $s=1$; or in other words, SPO (with single qubit intermediate Hamiltonians) achieves best-case performance even for the hard QUBO instances. 

We note that hard case 2 seems to undergo a finite-size analog of a phase transition around the minimum gap region under linear interpolation. This is indicated by an abrupt change in the ground state as $s_{\rm min}$ is crossed, as indicated by the rapid decay of fidelity between neighboring (in $s$) ground states \cite{Zanardi:2006fb,Zhou:2008cv}, which is shown in the inset of \ref{fig:hardpermy}(b). Such problem instances are known to be extremely difficult for AQO \cite{Amin:2009kf,Altshuler:2010ct,Knysh:2010un,Young:2010ct}.

\begin{figure}[tb]
\centering
 \psfrag{y1}[][][1]{Success Probability}
  \psfrag{x1}[][][1]{T}
 \psfrag{y1}[][][0.7]{Success Probability}
  \psfrag{x1}[][][0.6]{Evolution time $T$}
 \psfrag{aaaaa}[][][0.55]{$10^{-4}$}  \psfrag{bbbbb}[][][0.55]{$10^{-3}$}
  \psfrag{ccccc}[][][0.55]{$10^{-2}$}  \psfrag{ddddd}[][][0.55]{$10^{-1}$} \psfrag{eeeee}[][][0.55]{$1$}
  \psfrag{5}[][][0.55]{5}\psfrag{10}[][][0.55]{10} \psfrag{20}[][][0.55]{20}
  \psfrag{40}[][][0.55]{40} \psfrag{80}[][][0.55]{80} \psfrag{160}[][][0.55]{160} \psfrag{320}[][][0.5]{320} \psfrag{640}[][][0.5]{640} \psfrag{1280}[][][0.5]{1280}
\subfigure[~Hard case 1]{\includegraphics[width=0.2\textwidth]{\figurelib/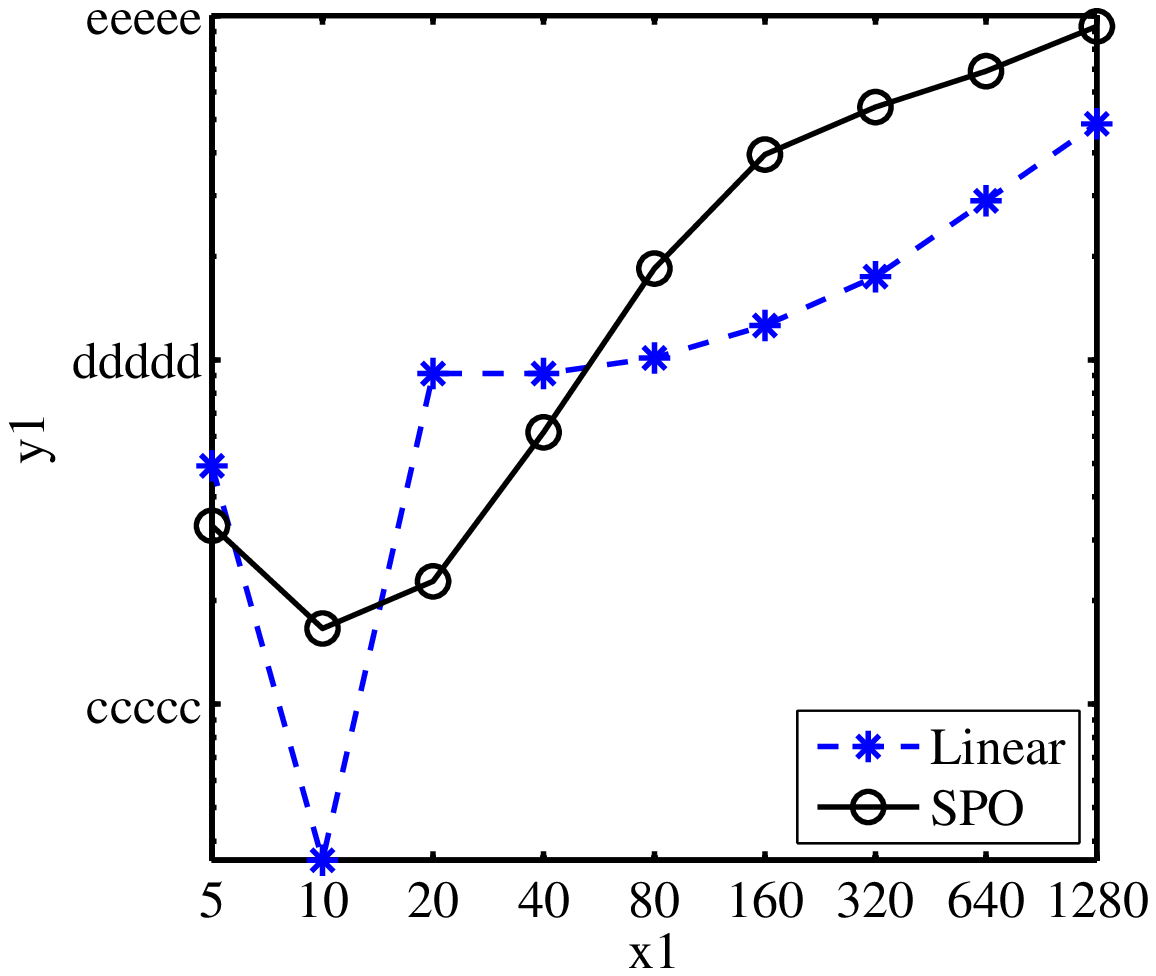}}
\subfigure[~Hard case 2]{\includegraphics[width=0.2\textwidth]{\figurelib/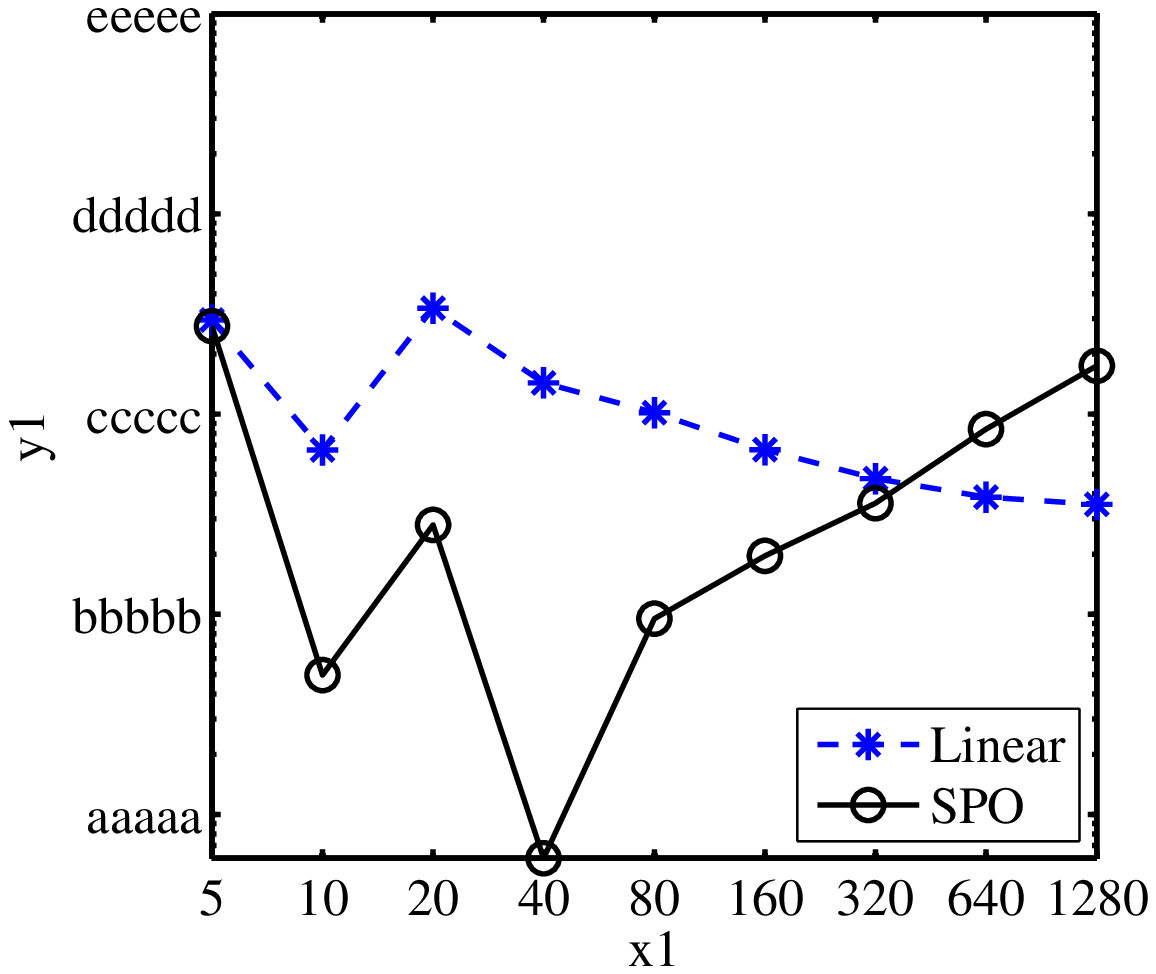}}
 \subfigure[~Hard case 3]{\includegraphics[width=0.2\textwidth]{\figurelib/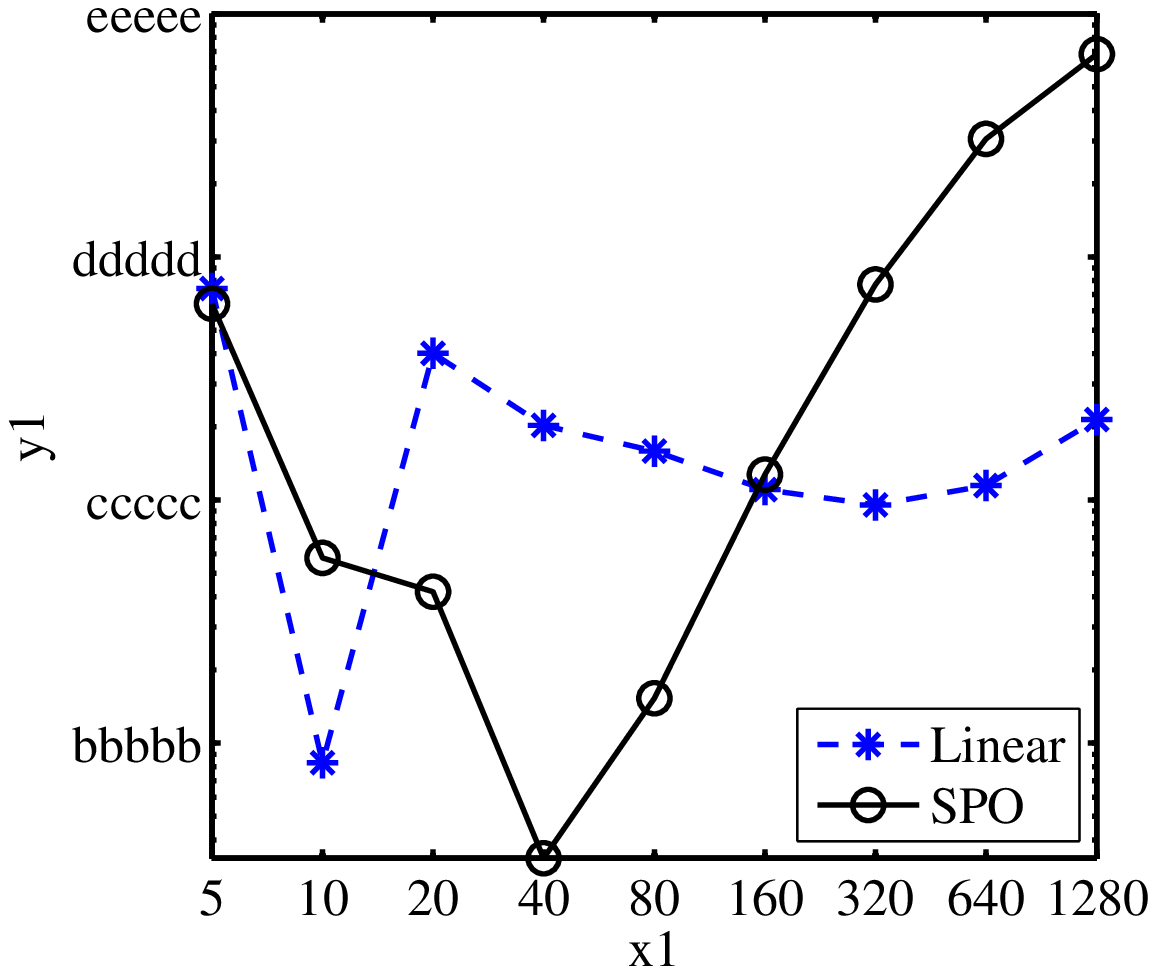}}
\caption{(Color online) Success probability as a function of running time $T$ for the three hardest instances of QUBO we found, with linear and SPO interpolations. For long $T$, SPO outperforms linear interpolation, even approaching $p_{\rm succ}=1$ for hard case 1 and 3. Note that the $x$-axis is not linear. The non-monotonic behavior of success probability with $T$ exhibited by the above curves is a well established phenomenon; non-adiabatic, fast sweeps may have better performance for hard problem instances \cite{Crosson:2014tg,Steiger:2015uk}.}
\label{fig:hard_succ}
\end{figure}

Furthermore, in Fig. \ref{fig:hard_succ} we show the probability of success as a function of $T$ for the three hardest instances when the linear interpolation or the optimized path is used. As in the easy QUBO cases analyzed above, for short $T$ the small increment in minimum gap does not directly translate to an increased probability of success. However, for large $T$ the optimized schedule does have a greater probability of success than linear interpolation. Notably, for hard case 2 the success probability cannot be significantly increased even for large $T$; \ie the difference in success probability between linear interpolation and SPO is not as significant for this case. We can understand this by noting that the phase transition, and associated avoided crossing, in this problem instance is not removed by the optimized adiabatic schedule (as evidenced in Fig. \ref{fig:hardpermy}(b) by the persistence of fidelity decay even with SPO), and thus the problem remains difficult for AQO. Also, we note that for hard cases 1 and 3, the success probability under linear interpolation can be dramatically increased by increasing the running time $T$. However, for fixed running time, especially for $T\sim 10$, these are difficult for AQO. 

We submit the results in this subsection and in subsection \ref{sec:optim} as numerical evidence that $\min_s \Delta(s) = \min \{\Delta(0), \Delta(1)\}$ can \emph{always} be achieved for QUBO with purely local intermediate Hamiltonians of the form \erf{eq:qubo_Hp}.

\section{Randomized intermediate Hamiltonians}
\label{sec:random}
The results in the previous section provide strong evidence that there exists at least one \emph{local} intermediate Hamiltonian for every QUBO problem that achieves best-case performance, \ie $\min_s \Delta(s) = \min \{\Delta(0), \Delta(1)\}$. However, finding this intermediate Hamiltonian is a difficult task in general. Therefore in this section we ask how well randomly chosen intermediate Hamiltonians can increase the minimum gap of the linear interpolation adiabatic evolution. Crosson \etal considered a similar question in Ref. \cite{Crosson:2014tg}, where they chose intermediate Hamiltonians of the form
\begin{equation}
\label{eq:farhi_random}
H^{\rm quad}_p(s) =s(1-s)\sum_{j=1}^n\sum_{\alpha=x,z}{\frac{c_{\alpha}^i}{\parallel c\parallel^2}}\sigma_{\alpha}^j,
\end{equation}
where the coefficients $c_{\alpha}^i$ are chosen randomly according to a Gaussian distribution. We have evaluated several forms for the random intermediate Hamiltonian, but will present results from sampling over intermediate Hamiltonians of the same form as \erf{eq:farhi_random} except for two differences: (i) whereas Crosson \etal considered $\sigma_{\alpha}^j$ that are one- \emph{and} two-qubit operators, we will use only local (one-qubit) perturbations, and (ii) we will sample $c_{\alpha}^i$ over the uniform distribution over $[-1,1]$, $(0,1]$, or $[-1,0)$. This form for the intermediate Hamiltonian has the benefit that it is smoothly varying as a function of $s$ for any random instance, and also the number of random variables needed is $2n$, as opposed to scaling with the discretization of the $s$ variable. This makes sampling over such random intermediate Hamiltonians much more feasible.

In this section we restrict our study to 10-qubit problems, and evaluate the change in the minimum gap achieved by the intermediate Hamiltonian:
\beq
\Omega \equiv \Delta^p_{\rm  min} - \Delta^0_{\rm min},
\eeq
where $\Delta^0_{\rm min}$ is the minimum gap for linear interpolation, and $\Delta^p_{\rm min}$ is the minimum gap for the Hamiltonian $H(s) = (1-s)H_0 + sH_1 + H^{\rm quad}_p(s)$. We will also investigate the change in success probability: $\Delta p = p_{\rm succ}^p - p_{\rm succ}^0$, where $p_{\rm succ}^p$ ($p_{\rm succ}^0$) is the success probability using the perturbing Hamiltonian (linear interpolation) with $T=10$.

\subsection{Average case behavior}
\begin{figure}[tb]
 \psfrag{x1}[][][0.8]{Instance number}
 \psfrag{x2}[][][0.8]{Instance number}
 \psfrag{y1}[][][0.8]{Random perturbation distribution}
 \psfrag{y2}[][][0.8]{Random perturbation distribution}
\centering
 \subfigure[~Distribution of $\Omega$.]{\includegraphics[width=0.4\textwidth]{\figurelib/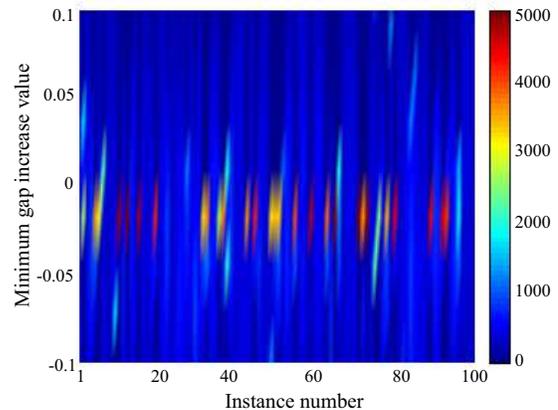}}
 \subfigure[~Distribution of $\Delta p_{\rm succ}$.]{\includegraphics[width=0.4\textwidth]{\figurelib/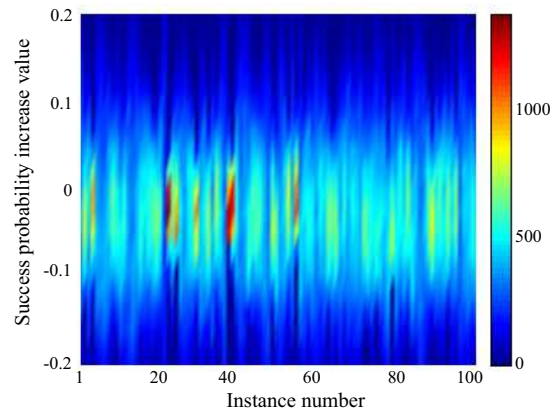}}
\caption{(Color online) The change in minimum gap (a) and success probability (b) for 100 QUBO problem instances under 5000 random perturbations of the form in \erf{eq:farhi_random}, when run for $T=10$. }
\label{fig:Gsp}
\end{figure}

First, we evaluate the effect of random intermediate Hamiltonians on randomly sampled QUBO instances. We generate $N_{\rm s}=100$ instances of QUBO problem Hamiltonians by sampling $h_i$ and $J_{ij}$ from a uniform distribution over $[-1,1]$, and for each one of these instances we assess the increase in minimum gap achieved by $N_{\rm r}=5000$ random intermediate Hamiltonians of the form \erf{eq:farhi_random}, with $c_{\alpha}^i$ sampled over the uniform distribution over $[-1,1]$. Fig. \ref{fig:Gsp} shows the distribution of $\Omega$ and $\Delta p_{\rm succ}$ over this random sampling of problems instances and perturbing Hamiltonians. If we average over the problems instances and the perturbing Hamiltonians, the average-case performance is summarized in Table \ref{tab:random_avg}. We see that the random perturbing intermediate Hamiltonian of the form in \erf{eq:farhi_random} does not increase performance with high probability, and on average seems to decrease the performance of AQC. 

\begin{table*}[tb]
  \centering
  \renewcommand\arraystretch{1.2}
    \begin{tabular}{c|c|c}
\hline
         & Change in minimum Gap, $\bar \Omega$ & Change in success probability, $\bar{\Delta }p_{\rm succ}$  \\
    \hline\hline
    Mean value and average percentiles & $-0.014828 ~(-0.0453,0.0195)$ &  $-0.016424 ~ (-0.0542, 0.0163)$ \\\hline
    \# increase & 7   & 12 \\
        \hline
    \end{tabular}%
    \caption{Change in minimum gap and success probability over 100 QUBO problem instances each with 5000 perturbing Hamiltonians. The first row shows the average change (averaged over problem instances and perturbing Hamiltonians) and average $35$th and $65$th percentiles in parentheses. For each problem instance these percentiles are computed over the 5000 perturbing Hamiltonians and then these percentiles are averaged over the 100 problem instances (this provides a measure of the variation in performance of random perturbing Hamiltonians, averaged over problem instances). The second row shows the number of problem instances (out of 100) where the average gap or average success probability increases.}
  \label{tab:random_avg}%
\end{table*}%

Next, we modify the character of the random perturbing intermediate Hamiltonian by restricting the domain over which the coefficients $c_{\alpha}^i$ in \erf{eq:farhi_random} are sampled. The four restrictions we consider are:
\begin{enumerate}
        \item All $c_{\alpha}^i$ positive;
        \item All $c_{\alpha}^i$ negative;
        \item $c_{\alpha}^i$ corresponding to local $\sigma_z$ terms are positive whereas the ones corresponding to local $\sigma_x$ terms are negative;
        \item $c_{\alpha}^i$ corresponding to local $\sigma_z$ terms are negative whereas the ones corresponding to local $\sigma_x$ terms are positive.
\end{enumerate}
For brevity, we do not present the full distribution of outcomes for each of these cases, but rather just summarize the average behavior in table \ref{tab:random_avg_restricted} (the averages are taken over the same 100 QUBO problem instances, and 1000 random perturbations in each case). We see that the average behavior with restricted $c_{\alpha}^i$ can be much better than when the domain of $c_{\alpha}^i$ is over $[-1,1]$. In particular, when all $c_{\alpha}^i$ are sampled from the positive interval $(0,1]$ the relative frequency of instances where the random intermediate Hamiltonian increased the success probability is greater than $0.5$, although the actual increase in success probability is modest.

\begin{table*}[htbp]
\renewcommand\arraystretch{1.2}
  \centering
      \begin{tabular}{c|c|c|c|c}
    \hline
    \textbf{$c_{\alpha}^i$ sampling restriction}  & \textbf{all $+$} & \textbf{all $-$} & \textbf{$\sigma_x:\;+,\;\sigma_z:\;-$} & \textbf{$\sigma_z:\;+,\;\sigma_x:\;-$} \\\hline\hline
    $\bar{\Omega}$ & \specialcell{$-2.2\times 10^{-5}$\\ $(-0.0144, 0.0160)$}& \specialcell{$-0.0213$\\$(-0.0431,-0.0118)$} & \specialcell{$0.0022$\\$(-0.0104, 0.0191)$} & \specialcell{$-0.0214$\\$(-0.0435,-0.0124)$} \\\hline
    $\bar{\Delta }p_{\rm succ}$ & \specialcell{$0.017451$\\$(0.0064,0.0404)$} & \specialcell{$-0.040459$\\$(-0.0714,-0.0373)$} & \specialcell{$0.006233$\\$(-0.0094,0.0264)$} & \specialcell{$-0.028102$\\$(-0.0538, -0.0214)$} \\\hline
     \# of increased minimum gap&56&27&44&41 \\
     \hline
\# of increased success probability & 68 & 22 & 55 & 31 \\
\hline
    \end{tabular}%
    \caption{Change in minimum gap ($\bar \Omega$) and success probability ($\bar{\Delta}p_{\rm succ}$) over 100 QUBO problem instances each with 1000 perturbing Hamiltonians, for each type of perturbing Hamiltonian presented in the main text. The first two rows show the average change (averaged over problem instances and perturbing Hamiltonians) and average $35$th and $65$th percentiles in parentheses. For each problem instance these percentiles are computed over the 1000 perturbing Hamiltonians and then these percentiles are averaged over the 100 problem instances (this provides a measure of the variation in performance of random perturbing Hamiltonians, averaged over problem instances). The last two rows show the number of problem instances (out of 100) where the average gap or average success probability (averaged over the 1000 random intermediate Hamiltonians) increases.}
  \label{tab:random_avg_restricted}%
\end{table*}%

\subsection{Hard QUBO instances}
\begin{figure}[t!]
\centering
   \psfrag{kaa}[][][0.57]{$\times 10^{-3}$}
   \psfrag{baa}[][][0.5]{}
 \psfrag{y1}[][][0.8]{Frequency}
  \psfrag{x1}[][][0.8]{Change in minimum gap}
   \psfrag{y2}[][][0.8]{Frequency}
  \psfrag{x2}[][][0.8]{Change in minimum gap}
   \psfrag{y3}[][][0.8]{Frequency}
  \psfrag{x3}[][][0.8]{Change in minimum gap}
\subfigure[~Hard case 1: mean 0.0040.]{\includegraphics[width=0.45\hsize]{\figurelib/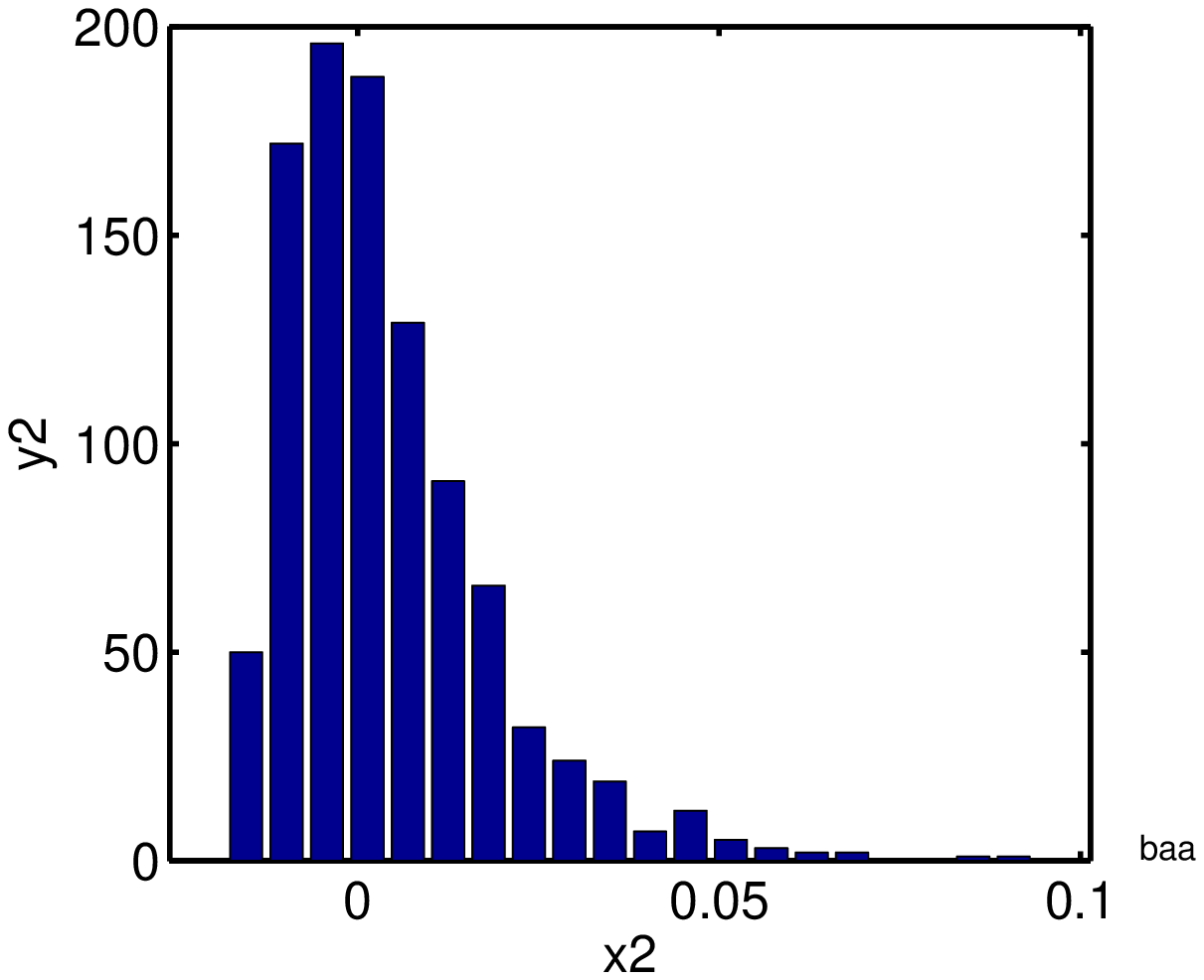}}
\subfigure[~Hard case 2: mean 0.0060.]{\includegraphics[width=0.45\hsize]{\figurelib/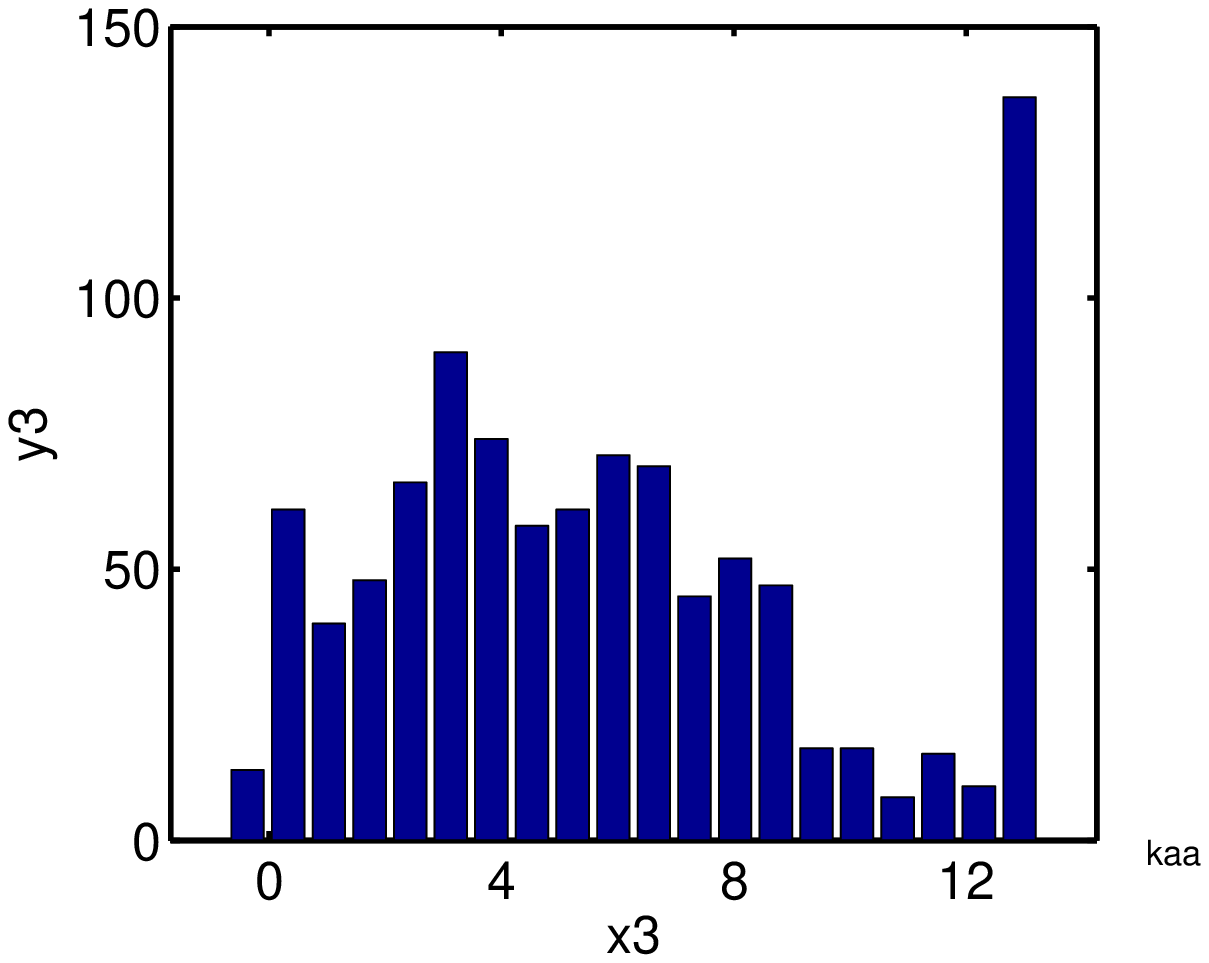}}
 \subfigure[~Hard case 3: mean $7.85\times 10^{-4}$.]{\includegraphics[width=0.45\hsize]{\figurelib/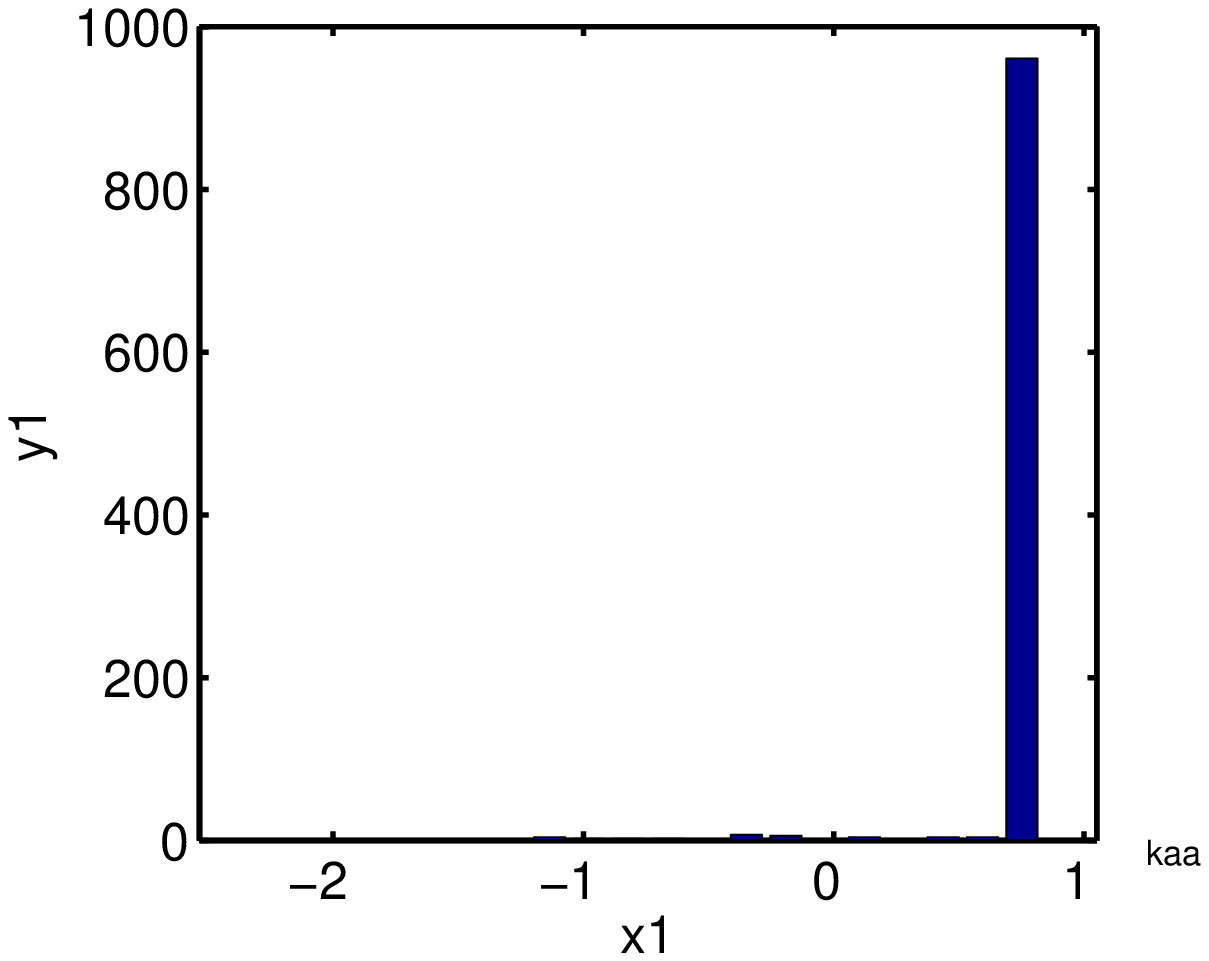}}
\caption{(Color online) Histograms of the change in minimum gap for the 3 hardest QUBO problems induced by random intermediate Hamiltonians. The caption for each subfigure shows the average gap change.}
\label{fig:gap_change_random}
\end{figure}

\begin{figure}[t!]
\centering
 \psfrag{y1}[][][0.8]{Frequency}
  \psfrag{x1}[][][0.8]{Change in success probability}
   \psfrag{y2}[][][0.8]{Frequency}
  \psfrag{x2}[][][0.8]{Change in success probability}
   \psfrag{y3}[][][0.8]{Frequency}
  \psfrag{x3}[][][0.8]{Change in success probability}
\subfigure[~Hard case 1: mean 0.0127.]{\includegraphics[width=0.2\textwidth]{\figurelib/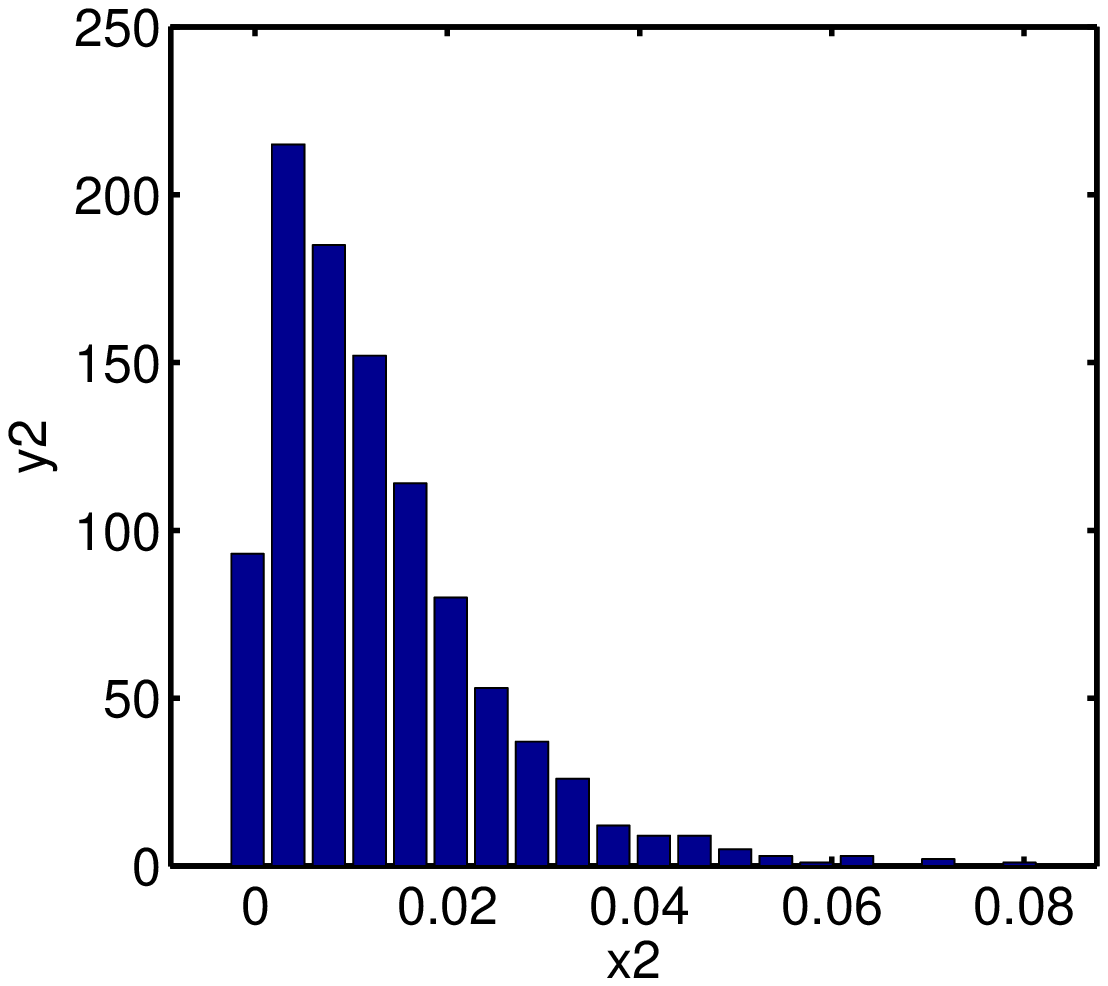}}
\subfigure[~Hard case 2: mean 0.0169.]{\includegraphics[width=0.2\textwidth]{\figurelib/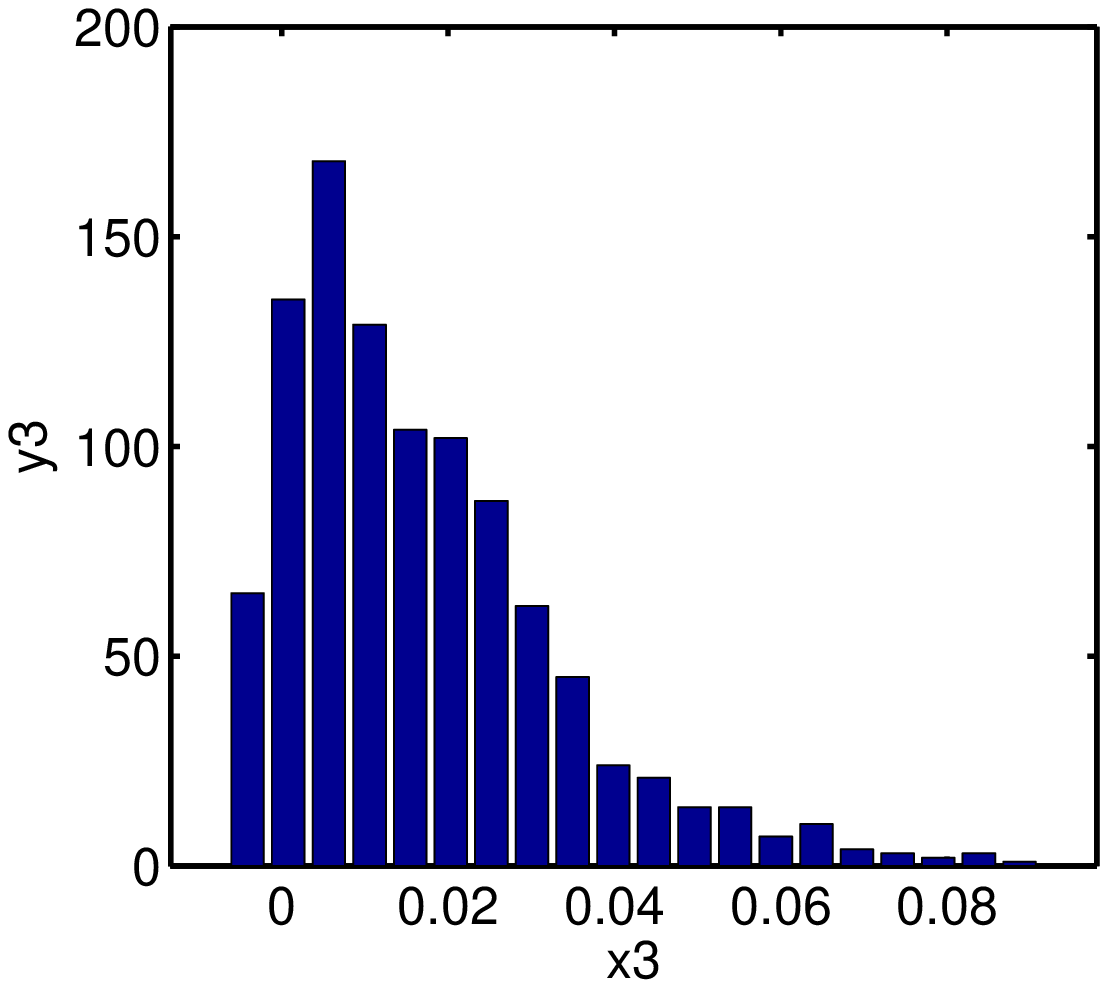}}
\subfigure[~Hard case 3: mean 0.0285.]{\includegraphics[width=0.2\textwidth]{\figurelib/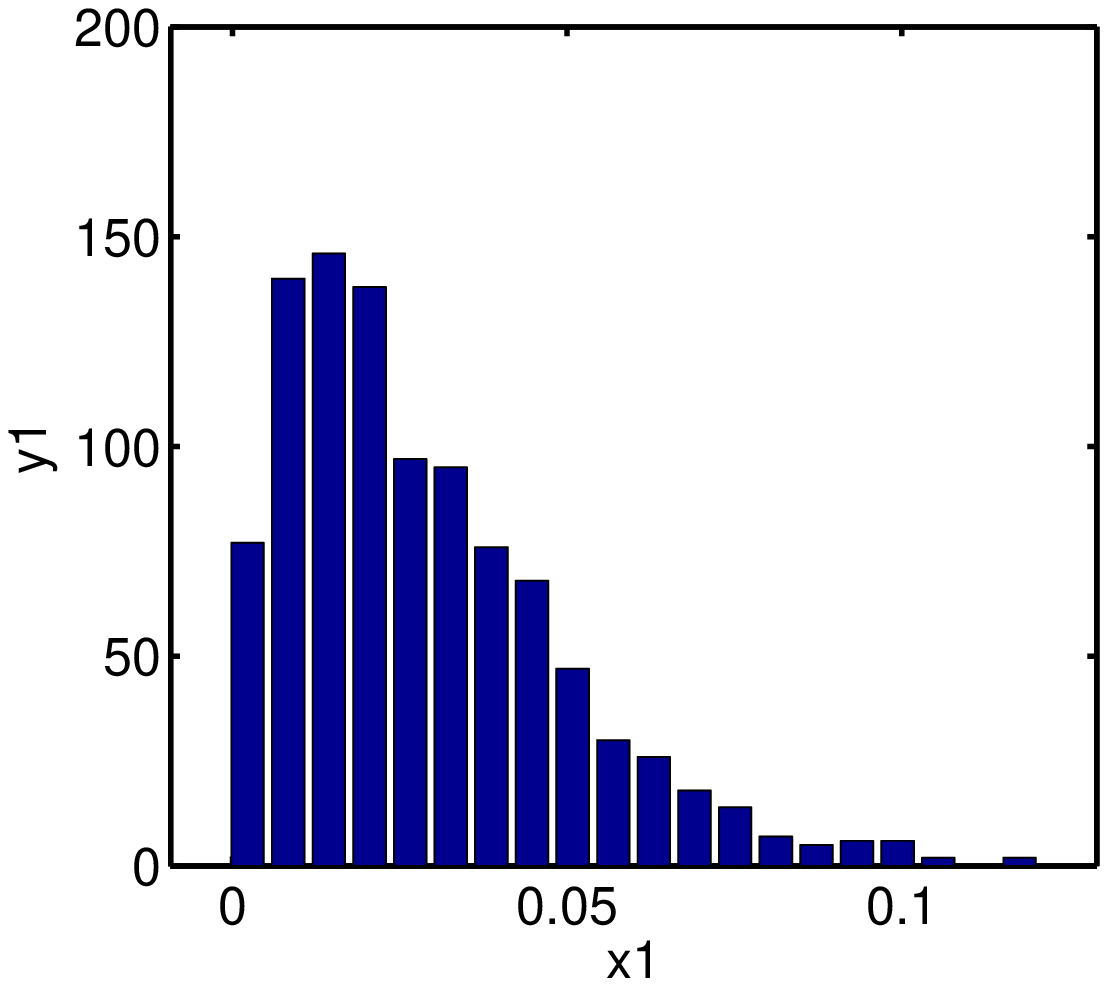}}
\caption{(Color online) Histograms of the change in success probability for the 3 hardest QUBO problems induced by random intermediate Hamiltonians. The caption for each subfigure shows the average change in success probability.}
\label{fig:sp_change_random}
\end{figure}

Now we return to the hardest instances of QUBO that we identified in section \ref{sec:hardest} and assess the effect of random intermediate Hamiltonians of the form \erf{eq:farhi_random} (with $c^i_{\alpha}$ sampled uniformly from the interval $(0,1]$) on solving them using adiabatic evolution. Figs. \ref{fig:gap_change_random} and \ref{fig:sp_change_random} show histograms of the changes in gap and in success probability, when 1000 random intermediate Hamiltonians of the form \erf{eq:farhi_random} are used to perturb the linear interpolation schedule. We see that the gap increases on average and moreover, increases in most cases for all three hard cases. Furthermore, the success probability increases with very high probability over the 1000 perturbing Hamiltonian samples for all three hard cases. This is in agreement with the observations in Ref. \cite{Crosson:2014tg}, although their intermediate Hamiltonians take a slightly different form. Therefore we have evidence that for these hardest instances of QUBO the strategy of introducing random local perturbations to the schedule path increases the performance with high probability, even though the degree of increase (in success probability) is very modest, at least for the form of intermediate Hamiltonians considered here.



\section{Discussion and Conclusions}
\label{sec:disc}
We studied the feasibility of increasing the success probability of AQO implementations using intermediate Hamiltonians to modify the adiabatic schedule. We began by examining the minimum gap during the interpolation between $H_0$ and $H_1$ as a proxy for success probability, and developed an approximative convex optimization formulation of the problem of finding intermediate Hamiltonians that achieve the maximum minimum gap, a problem we refer to to \emph{schedule path optimization} (SPO). Then by explicitly optimizing over intermediate Hamiltonians we collected convincing numerical evidence that there exists at least one purely local intermediate Hamiltonian that achieves the best-case performance (where the minimum gap is equal to the gap at one of the endpoints) for QUBO problem Hamiltonians. Next we evaluated the effect of SPO on the hardest QUBO problems, which are the ones with minimum gap very close to the endpoint ($s=1$), and concluded that although SPO can achieve an increase in gap and success probability, the improvements are modest.

Although we have evidence that SPO can achieve best-case performance, it is obviously not feasible for real AQC or AQO implementations. Therefore in section \ref{sec:random} we focused on random local intermediate Hamiltonians, sampled from a particular quadratic form of the schedule. We showed that if the random intermediate Hamiltonian is sampled correctly (with positive coefficients for each local term), the performance of AQO on randomly sampled QUBO problems can be improved with probability greater than $0.5$. This strategy also provides a very modest improvement in success probability for the hardest QUBO problem instances. More work is required to develop a heuristic strategy that improves performance for these very hard QUBO instances.

\begin{acknowledgments}
MS would like to thank Kevin Young, Constantin Brif, and Eddie Farhi for useful discussions. This work was supported by the Laboratory Directed Research and Development program at Sandia National Laboratories. Sandia is a multi-program laboratory managed and operated by Sandia Corporation, a wholly owned subsidiary of Lockheed Martin Corporation, for the United States Department of Energy's National Nuclear Security Administration under contract DE-AC04-94AL85000.  JZ acknowledges financial support from NSF China under Grant No. 61174086 and 61533012, and State Key Laboratory of Precision Spectroscopy, ECNU, China.
\end{acknowledgments}

\bibliographystyle{apsrev}
\bibliography{aqc_gap_engineering}

\end{document}